\documentstyle[12pt]{article}
\begin{document}
\def\beq{\begin{equation}}
\def\eeq{\end{equation}}
\def\bea{\begin{eqnarray}}
\def\eea{\end{eqnarray}}
\def\ve{\vert}
\def\vel{\left|}
\def\ver{\right|}
\def\nnb{\nonumber}
\def\ga{\left(}
\def\dr{\right)}
\def\aga{\left\{}
\def\adr{\right\}}
\def\rar{\rightarrow}
\def\nnb{\nonumber}
\def\la{\langle}
\def\ra{\rangle}
\def\lla{\left<}
\def\rra{\right>}
\def\ba{\begin{array}}
\def\ea{\end{array}}
\def\tep{$B \rar K \ell^+ \ell^-$}
\def\tepm{$B \rar K \mu^+ \mu^-$}
\def\tept{$B \rar K \tau^+ \tau^-$}
\def\ds{\displaystyle}



\newskip\humongous \humongous=0pt plus 1000pt minus 1000pt
\def\caja{\mathsurround=0pt}
\def\eqalign#1{\,\vcenter{\openup1\jot
\caja   \ialign{\strut \hfil$\displaystyle{##}$&$
\displaystyle{{}##}$\hfil\crcr#1\crcr}}\,}


\def\simlt{\stackrel{<}{{}_\sim}}
\def\simgt{\stackrel{>}{{}_\sim}}



\def\bos{\lower 0.5cm\hbox{{\vrule width 0pt height 1.2cm}}}
\def\boss{\lower 0.35cm\hbox{{\vrule width 0pt height 1.cm}}}
\def\aaa{\lower 0.cm\hbox{{\vrule width 0pt height .7cm}}}
\def\dol{\lower 0.4cm\hbox{{\vrule width 0pt height .5cm}}}


\title{ {\Large {\bf 
Model--independent analysis of the exclusive semileptonic
$B \rar \pi^+ \pi^- \ell^+ \ell^-$ decay  } } }

\author{\vspace{1cm}\\
{\small T. M. Aliev \thanks
{e-mail: taliev@metu.edu.tr}\,\,,
M. Savc{\i} \thanks
{e-mail: savci@metu.edu.tr}} \\
{\small Physics Department, Middle East Technical University} \\
{\small 06531 Ankara, Turkey} }
\date{}

\begin{titlepage}
\maketitle
\thispagestyle{empty}

\begin{abstract}
A model--independent analysis for the exclusive, rare $B \rar \pi^+ \pi^- \ell^+ \ell^-$ 
decay is presented. The dependence of the various physically measurable
asymmetries and CP violating asymmetry on the new Wilson coefficients is
studied in detail. It is observed that different measurable quantiles are
very sensitive to the different new Wilson coefficients, i.e., to the
existence of new physics beyond standard model. 
\end{abstract}

\end{titlepage}

\section{Introduction}
Rare $B$ meson decays, induced by flavor--changing neutral current (FCNC)
$b \rar s$ or $b \rar d$ transitions, occur at loop level in the standard
model (SM) and is a potential precision testing ground for the SM, as well as
attracting the theoretical interest as a promising tool for establishing
new physics beyond SM. New physics effects show themselves in rare $B$
meson decays in two different ways, namely, through the Wilson coefficients 
which could be distinctly different from their SM counterparts or through
the new structures in the effective Hamiltonian (see for example \cite{R1}--
\cite{R13}).
One of the main goal of the working $B$ factories and of the future hadron 
colliders is the study of the rare $B$ meson decays. Therefore 
theoretical and experimental 
investigation of the rare decays of 
$B$ mesons receive special attention. 
The aim of this paper to investigate the rare decays 
$\bar B \rar K^- \pi^+ \ell^+ \ell^-$ and 
$\bar B \rar \pi^+ \pi^- \ell^+ \ell^-$ when $K\pi$ and $\pi\pi$ systems are
the decay products of $K^\ast$ and $\rho$ mesons, in a model--independent
manner. 

The inclusive $b \rar s \ell^+ \ell^-$ and exclusive 
$B \rar K^\ast \ell^+ \ell^-$ decays were analyzed in a model--independent way
in \cite{R11} and \cite{R12}, respectively. The present work is an extension
of the previous studies of the $B \rar K^\ast(\rho) \ell^+ \ell^-$ decay. The
distribution in $\cos \theta_P$, where $\theta_P$ is the angle of the 
$K^-(\pi^-)$ in the $K^-\pi^+(\pi^-\pi^+)$ center of mass frame, and the
dependence on the azimuthal angle $\varphi$ between the $\ell^+ \ell^-$ and 
$K^-\pi^+$ or $\pi^-\pi^+$ planes, which does not exist in the 
$B \rar K^\ast(\rho) \ell^+ \ell^-$ decay, can provide additional new
information. This information is sensitive to the polarization state of the
vector meson $K^\ast(\rho)$ and therefore opens new possibility in probing
the structure of the effective Hamiltonian. Angular distributions 
and CP asymmetries in
the $\bar B \rar K^- \pi^+ \ell^+ \ell^-$ and
$\bar B \rar \pi^+ \pi^- \ell^+ \ell^-$ decays in the SM 
were studied in \cite{R14}. 

The paper is organized as follows. In section 2, we present the general,
model independent expression of the 
decay distribution for the $B \rar K \pi(\pi\pi) \ell^+ \ell^-$ decay, in
terms of the helicity amplitudes, including the non--zero lepton mass
effects. In section 3, we present our numerical analysis for different
angular distribution together with a  brief concluding remark of our
results.

\section{Theoretical background}

The matrix element for the $B \rar K^\ast(\rho) \ell^+ \ell^-$ decay
at quark level is described by the $b \rar f \ell^+ \ell^-~(f=s,d)$ transition.
Following the works \cite{R10}--\cite{R12}, the matrix elements of the
$b \rar f \ell^+ \ell^-$ decay can be written as the sum of the SM and new
physics contributions as
\bea
\label{matel}
{\cal M}_1 &=& \frac{G\alpha}{\sqrt{2} \pi}
 V_{tb}V_{tf}^\ast
\Bigg\{ C_{SL} \, \bar f i \sigma_{\mu\nu} \frac{q^\nu}{q^2}\, L \,b  
\, \bar \ell \gamma_\mu \ell + C_{BR}\, \bar f i \sigma_{\mu\nu}
\frac{q^\nu}{q^2} \,R\, b \, \bar \ell \gamma_\mu \ell \nnb \\
&+&C_{LL}^{tot}\, \bar f_L \gamma_\mu b_L \,\bar \ell_L \gamma^\mu \ell_L +
C_{LR}^{tot} \,\bar f_L \gamma_\mu b_L \, \bar \ell_R \gamma^\mu \ell_R +  
C_{RL} \,\bar f_R \gamma_\mu b_R \,\bar \ell_L \gamma^\mu \ell_L \nnb \\
&+&C_{RR} \,\bar f_R \gamma_\mu b_R \, \bar \ell_R \gamma^\mu \ell_R +
C_{LRLR} \, \bar f_L b_R \,\bar \ell_L \ell_R +
C_{RLLR} \,\bar f_R b_L \,\bar \ell_L \ell_R \nnb \\
&+&C_{LRRL} \,\bar f_L b_R \,\bar \ell_R \ell_L +
C_{RLRL} \,\bar f_R b_L \,\bar \ell_R \ell_L+
C_T\, \bar f \sigma_{\mu\nu} b \,\bar \ell \sigma^{\mu\nu}\ell \nnb \\
&+&i C_{TE}\,\epsilon^{\mu\nu\alpha\beta} \bar f \sigma_{\mu\nu} b \,
\bar \ell \sigma_{\alpha\beta} \ell  \Bigg\}~,
\eea
the first four terms contain the SM contribution too, while all the others
describe the new physics effects and
where 
\bea
L(R) = \frac{1-\gamma_5}{2} ~ \ga \frac{1+\gamma_5}{2} \dr \nnb~.
\eea
In further analysis we will neglect the tensor type interactions (terms
$\sim ~ C_T$ and $C_{TE}$ since physically measurable quantities are not
sensitive to the presence of tensor type interactions, as is shown in
\cite{R10,R11}). 

The matrix elements for the precess $B \rar P P^\prime \ell^+ \ell^-$ can
be obtained from the matrix element $B \rar V \ell^+ \ell^-~(V=K^\ast,~\rho)$ 
in the following way
\bea
\label{matpp}
{\cal M} = {\cal M}_{B \rar V \ell^+ \ell^-} \, \Pi(s_M) 
\la \pi^+(p_+) \pi^-(p_-) \ve V(p_V,\lambda) \ra~,
\eea  
where we assume that the resonance contribution of the intermediate vector
meson can be implemented by the Breit--Wigner form 
\bea
\label{PI}
\Pi(s_M) = \ds{\frac{\sqrt{m_V \Gamma_V /\pi}}{s_M -m_V^2 + i m_V
\Gamma_V} }~,
\eea
where $s_M = (p_+ + p_-)^2$ and $m_V$ and $\Gamma_V$ are the mass and
width of the vector meson ($K^\ast,~\rho$). For the decay part of the vector meson 
we use \cite{R15} 
\bea
\label{dprm}
\la P(p_+) P^\prime(p_-) \ve V(p_V,\lambda) \ra = \sqrt{BR} \,\,
Y_{\lambda_{max}}^{\lambda_i} (\theta_P,\varphi) ~,
\eea
where $Y_{\ell}^{m}(\theta_P,\varphi)$ are the $J=\ell$ spherical harmonics, and
the angles $\theta_P$ and $\varphi$ belong to those of the final $P$
meson in the vector meson's rest frame. The coupling of $V \rar P P^\prime$
decay is effectively taken into account by the branching ratios. 

From  (\ref{matpp}) we see that in order to calculate the matrix element 
$B \rar P P^\prime \ell^+ \ell^-$, the matrix element of the $B \rar V
\ell^+ \ell^-$  decay is needed in the first hand, for which the
matrix elements 
\bea
\label{roll}
\lla V \vel \bar f \gamma_\mu (1 \pm \gamma_5) b \ver B(p_B) \rra~,~
\lla V \vel \bar f i\sigma_{\mu\nu} (1 + \gamma_5) b \ver B(p_B)
\rra~~\mbox{\rm and}~~
\lla V \vel \bar f (1 \pm \gamma_5) b \ver B(p_B) \rra~,
\eea
need to be calculated, whose most general forms can be written as
\bea
\lefteqn{
\label{ilk}
\lla V(p_V,\varepsilon) \vel \bar f \gamma_\mu 
(1 \pm \gamma_5) b \ver B(p_B) \rra =} \nnb \\
&&- \epsilon_{\mu\nu\lambda\sigma} \varepsilon^{\ast\nu} p_{V}^\lambda q^\sigma
\frac{2 V(q^2)}{m_B+m_{V}} \pm i \varepsilon_\mu^\ast (m_B+m_{V})   
A_1(q^2) \mp i (p_B + p_{V})_\mu (\varepsilon^\ast q)
\frac{A_2(q^2)}{m_B+m_{V}} \nnb \\
&&\mp i q_\mu \frac{2 m_{V}}{q^2} (\varepsilon^\ast q)
\left[A_3(q^2)-A_0(q^2)\right]~, \nnb \\ \\
\lefteqn{
\label{iki}
\lla V(p_{V},\varepsilon) \vel \bar f i \sigma_{\mu\nu} q^\nu
(1 \pm \gamma_5) b \ver B(p_B) \rra =} \nnb \\
&&4 \epsilon_{\mu\nu\lambda\sigma} 
\varepsilon^{\ast\nu} p_{V}^\lambda q^\sigma
T_1(q^2) \pm 2 i \left[ \varepsilon_\mu^\ast (m_B^2-s_M) -
(p_B + p_{V})_\mu (\varepsilon^\ast q) \right] T_2(q^2) \nnb \\
&&\pm 2 i (\varepsilon^\ast q) \left[ q_\mu -
(p_B + p_{V})_\mu \frac{q^2}{m_B^2-s_M} \right] T_3(q^2)~,
\eea
where $q = p_B-p_V$ is the momentum transfer and $\varepsilon$ is the
polarization vector of the vector meson. 
In order to ensure finiteness of (\ref{ilk}) at $q^2=0$, 
we demand that $A_3(q^2=0) = A_0(q^2=0)$.  
The matrix element $\lla V \vel \bar f (1 \pm \gamma_5 ) b \ver B \rra$
can easily be calculated from Eq. (\ref{ilk}). For this aim it is enough to
contract both sides of Eq. (\ref{ilk}) with $q_\mu$ and use equation of
motion. Taking mass of the strange quark to be zero, it gives
\bea
\lefteqn{
\label{uc}
\lla V(p_V,\varepsilon) \vel \bar f (1 \pm \gamma_5) b \ver
B(p_B) \rra =
\frac{1}{m_b} \Big\{ \mp i (\varepsilon^\ast q) (m_B+m_V)
A_1(q^2)}\nnb \\
&&~~~~~~~~~~~~~\pm i (m_B^2-s_M) (\varepsilon^\ast q) \frac{A_2(q^2)}
{m_B+m_V}\pm 2 i m_V (\varepsilon^\ast q)
\left[A_3(q^2)-A_0(q^2)\right]~\Big\}~.
\eea
Furthermore, using the equation of motion, the form factor $A_3$ can
be expressed in terms of  the form factors $A_1$ and $A_2$ (see \cite{R16})
\bea
A_3(q^2) = \frac{m_B+m_V}{2 m_V} A_1(q^2) -
\frac{m_B-m_V}{2 m_V} A_2(q^2)~.
\eea
Using this relation, the matrix element (\ref{uc}) can be written in the
following form
\bea
\label{dort}
\lla V(p_V,\varepsilon) \vel \bar f (1 \pm \gamma_5) b \ver B(p_B) \rra
= \frac{1}{m_b} \Big\{\mp 2 i m_V (\varepsilon^\ast q) A_0(q^2)
\Big\}~.
\eea
As a result of the above considerations, the matrix element of the 
$B \rar V \ell^+ \ell^-$ decay can be determined straightforwardly
\bea
\lefteqn{
\label{had}
{\cal M}(B\rightarrow V \ell^{+}\ell^{-}) =
\frac{G \alpha}{4 \sqrt{2} \pi} V_{tb} V_{tf}^\ast }\nnb \\
&&\times \Bigg\{
\bar \ell \gamma_\mu(1-\gamma_5) \ell \, \Big[
-2 {\cal{V}}_{L_1}\epsilon_{\mu\nu\rho\sigma} \varepsilon^{\ast\nu}
p_V^V q^\sigma
 -i{\cal{V}}_{L_2}\varepsilon_\mu^\ast
+ i {\cal{V}}_{L_3}(\varepsilon^\ast q) (p_B+p_V)_\mu
+ i {\cal{V}}_{L_4}(\varepsilon^\ast q) q_\mu  \Big] \nnb \\
&&+ \bar \ell \gamma_\mu(1+\gamma_5) \ell \, \Big[
-2 {\cal{V}}_{R_1}\epsilon_{\mu\nu\rho\sigma} \varepsilon^{\ast\nu}
p_V^\rho q^\sigma
 -i{\cal{V}}_{R_2}\varepsilon_\mu^\ast    
+ i {\cal{V}}_{R_3}(\varepsilon^\ast q) (p_B+p_V)_\mu
+ i {\cal{V}}_{R_4}(\varepsilon^\ast q) q_\mu  \Big] \nnb \\
&&+\bar \ell (1-\gamma_5) \ell \Big[ i {\cal{S}}_{L}(\varepsilon^\ast
q)\Big]
+ \bar \ell (1+\gamma_5) \ell \Big[ i {\cal{S}}_{R}(\varepsilon^\ast
q)\Big]\Bigg\}~,
\eea
where ${\cal{V}}_{L_i}$ and ${\cal{V}}_{R_i}$ are the coefficients of left--
and right--handed
leptonic currents with vector structure, and
${\cal{S}}_{L,R}$ are the weights of
scalar leptonic currents with respective chirality, respectively, whose
explicit forms are given as
\bea
{\cal{V}}_{L_1} &=& (C_{LL}^{tot} + C_{RL}) \frac{V(q^2)}{m_B+m_V} -
2 (C_{BR}+C_{SL}) \frac{T_1}{q^2} ~, \nnb \\
{\cal{V}}_{L_2} &=& (C_{LL}^{tot} - C_{RL}) (m_B+m_V) A_1 - 2
(C_{BR}-C_{SL})
\frac{T_2}{q^2} (m_B^2-s_M) ~, \nnb \\
{\cal{V}}_{L_3} &=& \frac{C_{LL}^{tot} - C_{RL}}{m_B+m_V} A_2 - 2
(C_{BR}-C_{SL})
\frac{1}{q^2}  \left[ T_2 + \frac{q^2}{m_B^2-s_M} T_3 \right]~,
\nnb \\
{\cal{V}}_{L_4} &=& (C_{LL}^{tot} - C_{RL}) \frac{2 m_V}{q^2} (A_3-A_0)+
2 (C_{BR}-C_{SL}) \frac{T_3}{q^2} ~, \nnb \\
{\cal{V}}_{R_1} &=& {\cal{V}}_{L_1} ( C_{LL}^{tot} \rar C_{LR}^{tot}~,~~C_{RL} \rar
C_{RR})~,\nnb \\
{\cal{V}}_{R_2} &=& {\cal{V}}_{L_2} ( C_{LL}^{tot} \rar C_{LR}^{tot}~,~~C_{RL} \rar
C_{RR})~,\nnb \\
{\cal{V}}_{R_3} &=& {\cal{V}}_{L_3} ( C_{LL}^{tot} \rar C_{LR}^{tot}~,~~C_{RL} \rar
C_{RR})~,\nnb \\
{\cal{V}}_{R_4} &=& {\cal{V}}_{L_4} ( C_{LL}^{tot} \rar C_{LR}^{tot}~,~~C_{RL} \rar
C_{RR})~,\nnb \\
{\cal{S}}_{L} &=& - ( C_{LRRL} - C_{RLRL}) \ga \frac{2 m_V}{m_b} A_0
\dr~,\nnb \\
{\cal{S}}_{R} &=& - ( C_{LRLR} - C_{RLLR}) \ga \frac{2 m_V}{m_b} A_0
\dr~,\nnb   
\eea
In order to obtain the full helicity amplitude of the 
$B \rar P P^\prime \ell^+ \ell^-$ which follows from Eq. (\ref{matpp}), the
helicity amplitude of the $B \rar V \ell^+ \ell^-$ decay must be written,
which we denote as ${\cal M}_\lambda^{\lambda_\ell \bar \lambda_\ell}$
\bea
\label{hel}
{\cal M}_{\lambda_i}^{\lambda_\ell \bar \lambda_\ell} = 
\sum_{\lambda_{V^\ast}} \eta_{\lambda_{V^\ast}} 
L_{\lambda_{V^\ast}}^{\lambda_\ell \bar \lambda_\ell} 
H_{\lambda_{V^\ast}}^{\lambda_i}~,
\eea  
where
\bea
\eqalign{
\label{L}     
L_{\lambda_{V^\ast}}^{\lambda_\ell \bar \lambda_\ell}&=
\varepsilon_{V^\ast}^\mu 
\lla \ell^-(p_\ell,\lambda_\ell) \ell^+(p_\ell,\bar \lambda_\ell) 
\vel J^\ell_\mu \ver 0 \rra ~,\cr
H_{\lambda_{V^\ast}}^{\lambda_i \bar \lambda_i}&=  
\varepsilon_{V^\ast}^\mu 
\lla V(p_V,\lambda_i) 
\vel J_\mu^i \ver B(p_B) \rra ~,
}
\eea
where $\varepsilon_{V^\ast}$ is the polarization vector of the virtual
intermediate vector boson satisfying the relation 
\bea
-g^{\mu\nu} = \sum_{\lambda_{V^\ast}} \eta_{\lambda_{V^\ast}}
\varepsilon_{\lambda_{V^\ast}}^{\mu} 
\varepsilon_{\lambda_{V^\ast}}^{\ast\nu}~, \nnb
\eea
where the summation is over the helicities $\lambda_{V^\ast} = \pm 1,0,s$
of the virtual intermediate vector meson, with the metric defined as 
$\eta_\pm = \eta_0 = -\eta_s =1$ (for more detail see \cite{R17,R18}).
In Eq. (\ref{L}), $J_\mu^\ell$ and $J_\mu^i$ are the leptonic and hadronic
currents, respectively.  
Using Eqs. (\ref{had})--(\ref{L}) for the helicity amplitudes 
${\cal M}_{\lambda_i}^{\lambda_\ell \bar \lambda_\ell}$ we get the following
expressions
\bea
\eqalign{
\label{Ms}
{\cal M}^{++}_{\pm} &= \sin\theta A^{++}_{\pm} ~,\cr
{\cal M}^{+-}_{\pm} &= (-1\pm\cos\theta) A^{+-}_{\pm}~,\cr
{\cal M}^{-+}_{\pm} &= (1\pm\cos\theta) A^{-+}_{\pm}~,\cr
{\cal M}^{--}_{\pm} &= \sin\theta A^{--}_{\pm} ~,\cr
{\cal M}^{++}_{0}   &= \cos\theta A^{++}_{0} + B^{++}_{0} ~,\cr
{\cal M}^{+-}_{0}   &= \sin\theta  A^{+-}_{0}~,\cr
{\cal M}^{-+}_{0}   &= \sin\theta  A^{-+}_{0}~,\cr
{\cal M}^{--}_{0}   &= \cos\theta A^{--}_{0} + B^{--}_{0}~,
} 
\eea
where
\bea
\eqalign{
A^{++}_{\pm} &= \pm \sqrt{2} m_\ell \Big\{
( C_{LL}^{tot} + C_{LR}^{tot} ) H_\pm +
\frac{2}{q^2} \ga C_{BR} G_\pm + C_{SL} g_\pm \dr  + 
(C_{RR} + C_{RL}) h_\pm\Big\}~, \cr
A^{--}_{\pm} &= - A^{++}_{\pm} ~, \cr
A^{+-}_{\pm} &= \sqrt{\frac{q^2}{2}} \Big\{
\Big[ C_{LL}^{tot} (1-v) + C_{LR}^{tot} (1+v) \Big] H_\pm  
+ \Big[ C_{RL} (1-v) + C_{RR} (1+v) \Big] h_\pm \cr 
&+ \frac{2}{q^2} \ga  C_{BR} G_\pm + C_{SL} g_\pm \dr \Big\}~, \cr
A^{-+}_{\pm}   &= A^{+-}_{\pm} (v \rar -v)~, \cr
A^{++}_{0}   &= 2 m_\ell \Big[( C_{LL}^{tot} + C_{LR}^{tot} ) H_0 +
(C_{RL} + C_{RR}) h_0 + \frac{2}{q^2} \ga  C_{BR} G_0 + C_{SL} g_0
\dr\Big]~,\cr
A^{--}_{0}   &= - A^{++}_{0}~, \cr
B^{++}_{0}   &= - 2 m_\ell \Big\{
(C_{LR}^{tot} - C_{LL}^{tot}) H_S^0 +(C_{RR} - C_{RL}) h_S^0 \Big\} \cr
&-\frac{2}{m_b} q^2 \Big[ (1-v) ( C_{LRLR}- C_{RLLR}) -
(1+v) ( C_{LRRL}- C_{RLRL}) \Big] H_S^0 \Big\}~,\cr
B^{--}_{0}   &= B^{++}_{0} (v \rar -v)~, \cr
A^{+-}_{0}   &= -\sqrt{q^2} \Big\{
\Big[ C_{LL}^{tot} (1-v) + C_{RR} (1+v) \Big] H_0 +
\Big[ C_{RL} (1-v) + C_{RR} (1+v) \Big] h_0 \cr 
&+ \frac{2}{q^2} \ga C_{BR} G_0 + C_{SL} g_0 \dr \Big\}~, \cr
A^{-+}_{0}   &= A^{+-}_{0} (v \rar -v)~,
}
\eea
where superscripts denote helicities of the lepton and antilepton  and subscripts
correspond to the helicity of the vector meson (in our case $\rho$ or
$K^\ast$ meson),
respectively.

\bea
\label{Hpm}
\eqalign{
H_\pm &= \pm \lambda^{1/2}(m_B^2,s_M,q^2) \frac{V(q^2)}{m_B+m_V} +
(m_B+m_V) A_1(q^2)~,\cr
H_0 &= \frac{1}{2 \sqrt{s_M q^2}} \Bigg[
- (m_B^2-s_M-q^2) (m_B+m_V) A_1(q^2)  \cr
&+\lambda(m_B^2,s_M,q^2) \frac{A_2(q^2)}{m_B+m_V} \Bigg]~,\cr
H_S^0 &= \frac{\lambda^{1/2}(m_B^2,s_M,q^2)}{2 \sqrt{s_M q^2}} \Bigg[
- (m_B+m_V) A_1(q^2) +
\frac{A_2(q^2)}{m_B+m_V} (m_B^2-s_M) \cr 
&+ 2\sqrt{s_M} (A_3-A_0)\Bigg]~,\cr
G_\pm &= - 2 \Big[ \pm \lambda^{1/2}(m_B^2,s_M,q^2) T_1(q^2) +
(m_B^2-s_M)T_2(q^2) \Big]~,\cr
G_0 &= \frac{1}{\sqrt{s_M q^2}} 
\Bigg[ (m_B^2-s_M) (m_B^2-s_M-q^2) T_2(q^2)
- \lambda(m_B^2,s_M,q^2) \Bigg( T_2(q^2) \cr
&+ \frac{q^2}{m_B^2-s_M} T_3(q^2) \Bigg) \Bigg]~,\cr
h_\pm &= H_\pm(A_1 \rar -A_1,~A_2 \rar -A_2)~, \cr
\label{hS0}
h_S^0 &= H_S^0(A_1 \rar -A_1,~A_2 \rar -A_2)~,
}
\eea
where $\theta$ is the polar angle of positron in the rest frame of the
intermediate boson with respect to its helicity axis.
Note that we take $p_V^2 = s_M$, but not $m_V^2$, in order to take into account
$V$'s being a virtual particle which subsequently decays into 
$\pi^+ \pi^-$ or $K^- \pi^+$ pair. Remembering that the existing CLEO result \cite{R19}
for the $B \rar X_s \gamma$ and 
$B \rar K^\ast \gamma$ decays impose strong constraint on the parameter
space $C_{BR}$ and $C_{SL}$. For this reason here we assume that they are
equal to each other in the
SM. Hence we will take
\bea
C_{LL}^{tot} &=& C_9^{eff} - C_{10} + C_{LL}~, \nnb \\     
C_{LR}^{tot} &=& C_9^{eff} + C_{10} + C_{LR}~. \nnb
\eea
Using the expressions of the helicity amplitudes for the differential decay
rate width of the $B \rar V(\rar P P^\prime) \ell^+ \ell^-$ decay, we get
\bea
\label{dG}
\eqalign{
d\Gamma &=
\frac{3 G^2 \alpha^2}{2^{17} \pi^6 m_B^3 s_M q^2} \vel V_{tb} V_{tf}^\ast \ver^2
ds_M \, dq^2 \, d\cos\theta_P \, d\cos\theta \, d\varphi \cr
&\times \lambda^{1/2}(m_B^2,s_M,q^2) \lambda^{1/2}(s_M,m_P^2,m_{P^\prime}^2) 
\lambda^{1/2}(q^2,m_\ell^2,m_\ell^2)
\frac{m_V \Gamma_V /\pi}{(s_M-m_V^2)^2 + m_V^2 \Gamma_V^2} 
{\cal B}(V\rar P P^\prime) \cr
&\times\Bigg\{
2 \cos^2\theta_P \Big[\cos^2\theta \, N_1 + \sin^2\theta \, N_2 + 2 \cos\theta 
\,\mbox{\rm Re} [N_3] + N_4 \Big] \cr
&+\sin^2\theta_P \Big[ \sin^2\theta \,N_5 + \ga 1 + \cos^2\theta \dr N_6 + 2 \cos\theta
\,N_7 + 2 \sin (2\varphi) \sin^2\theta \,\mbox{\rm Im} [N_8] \cr
&- 2 \cos(2\varphi) \sin^2\theta \,\mbox{\rm Re} [N_8] \Big]
+ \sqrt{2} \sin(2\theta_P) \, \sin\theta \, \cos\varphi \,\mbox{\rm Re} 
\big[ \cos\theta \, N_9 + N_{10} \big] \cr
&-\sqrt{2} \sin(2\theta_P) \, \sin\theta \, \sin\varphi \,\mbox{\rm Im}
\big[ \cos\theta \, N_{11} + N_{12} \big] \Bigg\}~,
}
\eea
where
\bea
\label{Ns}
\eqalign{
N_1 &= \vel A_0^{++} \ver^2 + \vel A_0^{--} \ver^2~, \cr
N_2 &= \vel A_0^{+-} \ver^2 + \vel A_0^{-+} \ver^2~, \cr
N_3 &= A_0^{++}\ga  B_0^{++} \dr^\ast +  
        A_0^{--}\ga  B_0^{--} \dr^\ast ~, \cr
N_4 &= \vel B_0^{++} \ver^2 + \vel B_0^{--} \ver^2~, \cr
N_5 &= \vel A_+^{++} \ver^2 + \vel A_-^{++} \ver^2 +
        \vel A_+^{--} \ver^2 + \vel A_-^{--} \ver^2 ~, \cr
N_6 &= \vel A_+^{+-} \ver^2 + \vel A_-^{-+} \ver^2 +
        \vel A_-^{+-} \ver^2 + \vel A_+^{-+} \ver^2 ~, \cr
N_7 &= \vel A_-^{+-} \ver^2 + \vel A_+^{-+} \ver^2 
        - \vel A_+^{+-} \ver^2 - \vel A_-^{-+} \ver^2 ~, \cr
N_8 &= A_+^{++}\ga  A_-^{++} \dr^\ast + 
        A_+^{+-}\ga  A_-^{+-} \dr^\ast + 
        A_+^{-+}\ga  A_-^{-+} \dr^\ast + 
        A_+^{--}\ga  A_-^{--} \dr^\ast ~, \cr
N_{9} &= A_0^{++}\ga  A_-^{++} - A_+^{++} \dr^\ast - 
           A_0^{+-}\ga  A_-^{+-} + A_+^{+-} \dr^\ast - 
           A_0^{-+}\ga  A_-^{-+} + A_+^{-+} \dr^\ast \cr
&+ A_0^{--}\ga  A_-^{--} - A_+^{--} \dr^\ast ~, \cr
N_{10} &= B_0^{++}\ga  A_-^{++} - A_+^{++} \dr^\ast + 
           A_0^{+-}\ga  - A_-^{+-} + A_+^{+-} \dr^\ast + 
           A_0^{-+}\ga  A_-^{-+} - A_+^{-+} \dr^\ast \cr   
&+ B_0^{--}\ga  A_-^{--} - A_+^{--} \dr^\ast ~, \cr
N_{11} &= N_{9} \ga A_+^{++} \rar - A_+^{++},~A_+^{+-} \rar - A_+^{+-},~
           A_+^{-+} \rar - A_+^{-+},~A_+^{--} \rar - A_+^{--} \dr ~, \cr
N_{12} &= N_{10} \ga A_+^{++} \rar - A_+^{++},~A_+^{+-} \rar - A_+^{+-},~
           A_+^{-+} \rar - A_+^{-+},~A_+^{--} \rar - A_+^{--} \dr ~,
}
\eea
where $\theta_P$ is the polar angle of the pseudoscalar $P$ meson momentum
in the rest frame of the vector meson, with respect to the helicity axis, i.e.,
the outgoing direction of $V$ meson, and $\varphi$ is the azimuthal angle
between the planes of the two decays $V \rar P P^\prime$ and 
$V^\ast \rar \ell^+ \ell^-$. Kinematically allowed region of the variables
are given as
\bea
\label{kin}\eqalign{
\ga m_P + m_{P^\prime} \dr^2 &\le s_M \le \ga m_B - 2 m_\ell \dr^2 ~, \cr
4 m_\ell^2 &\le q^2 \le \ga m_B - \sqrt{s_M} \dr^2 ~, \cr
-1 &\le \cos\theta_P \le 1 ~, \cr
-1 &\le \cos\theta \le 1 ~, \cr
0 &\le \varphi \le 2 \pi~.}
\eea
We note that, in further analysis the narrow--width approximation for $V$
meson will be used, i.e., 
\bea
\label{lim}
\lim_{\Gamma_V \rar 0}\,
\ds{\frac{\Gamma_V m_V/\pi}{(s_M -m_V^2)^2 + m_V^2 \Gamma_V^2} } = 
\delta(s_M-m_V^2)~,\nnb
\eea
by means of which integration of Eq. (\ref{dG}) over $s_M$ can easily be
carried and the differential decay rate with respect to dilepton
mass $q^2$, azimuthal angle $\varphi$, polar angles $\theta_P$ and $\theta$
can be written as
\bea
\label{dG2}
\eqalign{
d\Gamma &=
\frac{3 G^2 \alpha^2}{2^{17} \pi^6 m_B^3 m_V^2q^2} \vel V_{tb} V_{tf}^\ast \ver^2
{\cal B}(V\rar P P^\prime)\, 
dq^2\, d\cos\theta_P\, d\cos\theta \,d\varphi \cr
&\times \lambda^{1/2}(m_B^2,m_V^2,q^2) \lambda^{1/2}(m_V^2,m_P^2,m_{P^\prime}^2) 
\lambda^{1/2}(q^2,m_\ell^2,m_\ell^2) \cr
&\times\Bigg\{
2 \cos^2\theta_P \Big[\cos^2\theta \, N_1 + \sin^2\theta \, N_2 + 2 \cos\theta 
\,\mbox{\rm Re} [N_3] + N_4 \Big] \cr
&+\sin^2\theta_P \Big[ \sin^2\theta \,N_5 +\ga 1 + \cos^2\theta \dr N_6 + 2 \cos\theta
\,N_7 + 2 \sin (2\varphi) \sin^2\theta \,\mbox{\rm Im} [N_8] \cr
&- 2 \cos(2\varphi) \sin^2\theta \,\mbox{\rm Re} [N_8] \Big]
+ \sqrt{2} \sin(2\theta_P) \, \sin\theta \, \cos\varphi \,\mbox{\rm Re} 
\big[ \cos\theta \, N_{9} + N_{10} \big] \cr
&-\sqrt{2} \sin(2\theta_P) \, \sin\theta \, \sin\varphi \,\mbox{\rm Im}
\big[ \cos\theta \, N_{11} + N_{12} \big] \Bigg\}~,
}
\eea
for which we will use the experimental results for the $V \rar P P^\prime$ namely, 
${\cal B}(\rho \rar \pi^+ \pi^-) = {\cal B}(K^\ast \rar K \pi) = 1$.
It should be noted here that in addition to the variables that exist in 
$B \rar V \ell^+ \ell^-$ decay, there appears a new variable $\theta_P$, and
contrary to the $B \rar V \ell^+ \ell^-$ case, the dependence of the cascade decay  
$B \rar V ( \rar P P^\prime) \ell^+ \ell^-$ on the azimuthal angle $\varphi$ 
is not trivial. Incidentally, we would like to remind the reader that, 
if Eq. (\ref{dG2}) is integrated over $\theta_P$ and $\varphi$, 
the differential decay 
rate for the $B \rar V \ell^+ \ell^-$ decay is obtained. The model independent 
analysis of the  $B \rar K^\ast \ell^+ \ell^-$ decay is presented in
\cite{R12}, in which the dependence of the experimentally measured 
quantities, such as 
branching ratio, forward--backward asymmetry and longitudinal polarization 
of the final lepton and the ratio $\Gamma_L/\Gamma_T$ of the decay widths
when $K^\ast$ meson is longitudinally and transversally polarized, on the
new Wilson coefficients, are systematically studied. As has been noted
already, our main goal in this work is to investigate the dependence of such
angular distributions on the new Wilson coefficients in the
$B \rar V(V\rar P P^\prime) \ell^+ \ell^-$ decay, which does not exist in
the $B \rar V \ell^+ \ell^-$ decay and not studied in \cite{R12}.

It can easily be seen from Eq. (\ref{dG2}) that the cascade decay 
$B \rar V ( \rar P P^\prime) \ell^+ \ell^-$ has a rich angular structure.
Therefore, in the light of this observation, a thorough investigation of 
different distributions will prove useful in separating various angular
coefficients experimentally. Along the lines as suggested by \cite{R18}, we
adopt two different strategies in further analysis of the problem under
consideration, namely investigation of the various individual angular
distributions and asymmetries and their relation to the new Wilson
coefficients.  

For the purpose of studying single angle distributions, 
we integrate Eq. (\ref{dG2}) over $q^2$, $\theta$ and 
$\varphi$, which takes the form
\bea
\label{dGP}
\ds{\frac{d \Gamma}{d \cos\theta_P}} \sim
2 \cos^2\theta_P \Bigg[ \frac{2}{3}\, \widetilde N_1 +
\frac{4}{3} \, \widetilde N_2 + 2 \widetilde N_4\Bigg] +
\sin^2\theta_P \Bigg[\frac{4}{3} \, \widetilde N_5 +
+ \frac{8}{3}\, \widetilde N_6 \Bigg] ~,
\eea
where we introduce the notation $\widetilde N_i = \int N_i \, d q^2$.
Defining an asymmetry parameter $\alpha_{\theta_P}$, from the angular 
distribution $W(\cos\theta_P)= 1 + \alpha_{\theta_P} \cos^2\theta_P$ we
get
\bea
\label{alp}
\alpha_{\theta_P} = \ds{\frac{\widetilde N_1 + 2 \widetilde N_2 +
3 \widetilde N_4}{\widetilde N_5 + 2 \widetilde N_6
}} - 1~.
\eea
Integrating Eq. (\ref{dG2}) over $q^2$, $\theta_P$ and $\varphi$, for
the polar $\cos\theta$ distribution we get
\bea
\label{dGt}
\eqalign{
\ds{\frac{d \Gamma}{d \cos\theta}} &\sim \frac{4}{3} \Big[
\cos^2\theta \, \widetilde N_1 + \sin^2\theta \, \widetilde N_2 +
2 \cos\theta \, \mbox{\rm Re} [\widetilde N_3] + \widetilde N_4 \Big] \cr
&+ \frac{4}{3} \Big[ \sin^2\theta \, \widetilde N_5 + 
\ga 1 + \cos^2\theta \dr \widetilde N_6 + 2 \cos\theta \, \widetilde N_7 
\Big]~,\nnb
}
\eea
from which one can write the angular distribution 
$W(\cos\theta)= 1 + \alpha_\theta \cos\theta + 
\beta_\theta \cos^2\theta$, with the asymmetry parameters
$\alpha_{\theta}$ and $\beta_{\theta}$ being defined as
\bea
\label{alt}
\eqalign{
\alpha_\theta &= \ds{\frac{2 \mbox{\rm Re} [\widetilde N_3] + 
2 \widetilde N_7}{\widetilde N_2 + \widetilde N_4 +
\widetilde N_5 + \widetilde N_6}}~,\cr
\label{bet}
\beta_\theta &= \ds{\frac{\widetilde N_1 - 
\widetilde N_2 - \widetilde N_5 + \widetilde N_6}
{\widetilde N_2 + \widetilde N_4 + 
\widetilde N_5 + \widetilde N_6}}~.
}
\eea
Finally we consider the azimuthal angle $\varphi$ distribution, which is
obtained by integrating Eq. (\ref{dG2}) over the parameters $q^2$, $\theta_P$
and $\theta$ to yield
\bea
\label{dGph}
\eqalign{
\ds{\frac{d \Gamma}{d \varphi}} &\sim \frac{4}{3} \Bigg[
\frac{2}{3}\, \widetilde N_1 + \frac{4}{3} \, \widetilde N_2 +
2 \widetilde N_4 \Bigg] \cr
&+ \frac{4}{3} \Bigg[ \frac{4}{3} \, \widetilde N_5 + 
\frac{8}{3} \, \widetilde N_6 + 
\frac{8}{3} \sin(2\varphi) \, \mbox{\rm Im} [\widetilde N_9] -
\frac{8}{3} \cos(2\varphi) \, \mbox{\rm Re} [\widetilde N_9] \Bigg]~,\nnb
}
\eea
and the azimuthal angle $\varphi$ asymmetry parameters $\alpha_\varphi$ and 
$\alpha_\varphi$ can be extracted from 
$W(\varphi)= 1 + \alpha_\varphi \sin(2\varphi)  + 
\beta_\varphi \cos(2\varphi)$ to give  
\bea
\label{alpph} 
\eqalign{
\alpha_\varphi &= \ds{\frac{4 \mbox{\rm Im} [\widetilde N_9]}
{\widetilde N_1 + 2 \widetilde N_2 + 
3 \widetilde N_4 + 2 \widetilde N_5 + 4 \widetilde N_6 }}~,\cr
\label{bep}
\beta_\varphi &= \ds{\frac{- 4 \mbox{\rm Re} [\widetilde N_9]}
{\widetilde N_1 + 2 \widetilde N_2 +
3 \widetilde N_4 + 2 \widetilde N_5 + 4 \widetilde N_6 }}~.
}
\eea

A second strategy for separating various angular coefficients
experimentally, is to define suitable asymmetry ratios that project out the
partial rates from Eq. (\ref{dG2}). For this purpose we consider the
following asymmetries (see also \cite{R18})
\bea
\label{Aph}
A_\varphi &=& \ds{
\frac{d\Gamma(\varphi) - d\Gamma(\varphi+\pi/2) + d\Gamma(\varphi+\pi) -
d\Gamma(\varphi+3\pi/2)}
{d\Gamma(\varphi) + d\Gamma(\varphi+\pi/2) + d\Gamma(\varphi+\pi) +
d\Gamma(\varphi+3\pi/2)}
}~, \\ \nnb \\
&&- \frac{\pi}{4} \le \varphi \le \frac{\pi}{4}~, \nnb \\ \nnb \\ 
\label{A1}
A_1 &=& \ds{\frac{N}{D}}~,
\eea
where
\bea
N &=& 
d\Gamma(\theta,\theta_P,\varphi) -
d\Gamma(\theta,\theta_P,\varphi+\pi) - 
d\Gamma(\theta,\pi-\theta_P,\varphi) +
d\Gamma(\theta,\pi-\theta_P,\varphi+\pi) \nnb \\
&-&
d\Gamma(\pi-\theta,\theta_P,\varphi) +
d\Gamma(\pi-\theta,\theta_P,\varphi+\pi) + 
d\Gamma(\pi-\theta,\pi-\theta_P,\varphi) \nnb \\
&-&
d\Gamma(\pi-\theta,\pi-\theta_P,\varphi+\pi)~, \nnb \\ \nnb \\
&&\phantom{-}0 \le \theta_P \le \pi/2~, \nnb \\
&&\phantom{-}\frac{\pi}{2} \le \theta \le \pi~, \nnb \\
&&-\frac{\pi}{2} \le \varphi \le \frac{\pi}{2}~,\nnb 
\eea
and the denominator $D$ is given by the same expression, 
with plus signs everywhere,
\bea
\label{A2}
A_2 &=& \ds{
\frac{d\Gamma(\theta_P,\varphi) - d\Gamma(\theta_P,\varphi+\pi) 
- d\Gamma(\pi-\theta_P,\varphi) + d\Gamma(\pi-\theta_P,\varphi+\pi)}
{d\Gamma(\theta_P,\varphi) + d\Gamma(\theta_P,\varphi+\pi) 
+ d\Gamma(\pi-\theta_P,\varphi) + d\Gamma(\pi-\theta_P,\varphi+\pi)}
}~, \\ \nnb \\      
&&\phantom{-}0 \le \theta_P \le \pi/2~, \nnb \\
&&-\frac{\pi}{2} \le \varphi \le \pi/2~. \nnb 
\eea
In expressions (\ref{Aph})--(\ref{A2}) the angles that do not appear in the
arguments of the differential rate $d\Gamma$ have been integrated out over
their physical ranges $(0 \le \theta,~\theta_P \le \pi,~0 \le \varphi \le 2
\pi)$. Integrating over all variables, we are left with the expressions 
$\widetilde A_\varphi,~\widetilde A_1,~\widetilde A_2$, which depend only on
the Wilson coefficients, as follows (here $~\widetilde{}~$ in the notation refers to
integration's being performed over all variables)
\bea
\label{Apht}
\widetilde A_\varphi &=& \ds{
\frac{- 8 \mbox{\rm Re} [\widetilde N_9]}
{\pi \ga \widetilde N_1 + 2 \widetilde N_2 +
3 \widetilde N_4 + 2 \widetilde N_5 + 4\widetilde N_6 \dr}}~,\\ \nnb \\
\label{A1t} 
\widetilde A_1 &=& \ds{
\frac{- \sqrt{2} \mbox{\rm Re} [\widetilde N_{10}]}
{\pi \ga \widetilde N_1 + 2 \widetilde N_2 +
3 \widetilde N_4 + 2 \widetilde N_5 + 4\widetilde N_6 \dr}}~,\\ \nnb \\
\label{A2t} 
\widetilde A_2 &=& \ds{
\frac{3 \mbox{\rm Re} [\widetilde N_{11}]}
{\sqrt{2} \ga \widetilde N_1 + 2 \widetilde N_2 +
3 \widetilde N_4 + 2 \widetilde N_5 + 4\widetilde N_6 \dr}}~.
\eea
Before proceeding further, we would like to consider the CP--violating
observables that can be constructed by combining the information on $\bar B$
and $B$ decays, namely, $\bar B \rar P \bar P^\prime \ell^+ \ell^-$ and
$B \rar \bar P P^\prime \ell^+ \ell^-$, and define CP--odd asymmetry in
the following way
\bea
\widetilde A_{CP} \equiv \ds{\frac{\Gamma - \bar \Gamma}
{\Gamma + \bar \Gamma} }~,
\eea 
where $\Gamma$ and $\bar \Gamma$ are the decay widths of the 
$\bar B \rar P \bar P^\prime \ell^+ \ell^-$ and $B \rar \bar P P^\prime
\ell^+ \ell^-$ processes, respectively. Explicit form of $\Gamma$ can easily
be obtained from Eq. (\ref{dG2}), by performing integration over the
variables $q^2,~\theta,~\theta_P,$ and $ \varphi$. The decay width for the
conjugate process can again be obtained from Eq. (\ref{dG2}) by making the
replacement $N_i \rar \bar N_i$, where $\bar N_i$ are the functions for the
conjugate processes. It should be noted that we consider a case in which all
new Wilson coefficients are real. Furthermore, form factors which enter into 
Eq. (\ref{Hpm}) are computed in framework of light cone QCD 
sum rules method \cite{R20}--\cite{R22} and this nonperturbative approach  
predicts that all form factors are real as well. In other words, there is no
any new source for CP violation other than that are present in the SM. As is
well known, for the $B \rar \rho \ell^+ \ell^-$ decay in the SM only the coefficient 
$C_9^{eff}$ contains both a weak
phase $\varphi_W$ (associated with the imaginary part of the CKM matrix
element) and a strong phase $\delta_S$ (attributed to the imaginary parts of
the $c \bar c$ and $b \bar b$ loops). As a result of these considerations,
it follows then that the decay width for the conjugate 
$B \rar \bar P P^\prime\ell^+ \ell^-$ process can be obtained from 
$\bar B \rar P \bar P^\prime \ell^+ \ell^-$ channel by the replacements
$\varphi_W \rar - \varphi_W$ and $\delta_S \rar \delta_S$.    

\section{Numerical analysis}

In this section we present our numerical results for the asymmetries
$A_{CP},~\alpha_{\theta_P},~\alpha_\theta,~\beta_\theta$, $\beta_\varphi~,
A_\varphi~,\alpha_\varphi~,\widetilde A_1~,\widetilde A_\varphi$ and $\widetilde 
A_2$ for the
exclusive rare $B \rar \pi^+ \pi^- \ell^+ \ell^-$ decay only. 
We take hadronic form factors from Table I
and the Wilson coefficients
from Table II. The values of the main input parameters used in our analysis
are: $m_b=4.8~GeV,~m_c=1.35~GeV,~m_\rho=0.77~GeV,~m_\tau=1.78~GeV,~m_\mu=0.105~GeV$
and $m_B=5.28~GeV$. Here we note that the results for $B \rar K \pi \ell^+ \ell^-$ 
can be obtained from $B \rar \pi \pi \ell^+ \ell^-$ by replacing 
$V_{td}$ with $V_{ts}$, $B \rar \rho$ transition form factors with $B \rar
K$ transition form factors, and $m_\pi$ with $m_K$.

In the present work we choose
light cone QCD sum rules method predictions for the form factors. In our
numerical analysis 
we will use the results of the work \cite{R21,R22} in which the form 
factors are described by a three--parameter fit where the radiative 
corrections up to leading twist contribution and
SU(3)--breaking effects are taken into account.
The $q^{2}$--dependence of the form factors, which appears in our analysis could
be parametrized as
\bea
\label{formfac}
F(s) = \frac{F(0)}{1-a_F\,s + b_F\, s^{2}}~, \nnb
\eea
where $s = q^2/m_B^2$ is the dilepton invariant mass in units of 
$B$ meson mass, and the parameters $F(0)$, $a_F$ and $b_F$ are 
listed in Table 1 for each
form factor.
\begin{table}[h]
\renewcommand{\arraystretch}{1.5}
\addtolength{\arraycolsep}{3pt}
$$
\begin{array}{|l|ccc|}  
\hline
& F(0) & a_F & b_F \\ \hline
A_0^{B \rar \rho} &\phantom{-}0.372 & 1.40 & \phantom{-} 0.437 \\
A_1^{B \rar \rho} &\phantom{-}0.261 & 0.29 & -0.415 \\ 
A_2^{B \rar \rho} &\phantom{-}0.223 & 0.93 & -0.092 \\
V^{B \rar \rho}   &\phantom{-}0.338 & 1.37 & \phantom{-} 0.315\\
T_1^{B \rar \rho} &\phantom{-}0.143 & 1.41 & \phantom{-} 0.361\\
T_2^{B \rar \rho} &\phantom{-}0.143 & 0.28 & -0.500 \\
T_3^{B \rar \rho} &\phantom{-}0.101 & 1.06 & -0.076 \\ \hline
\end{array}
$$
\caption{The form factors for $B\rightarrow \rho \ell^{+}\ell^{-}$
in a three--parameter fit.}
\renewcommand{\arraystretch}{1}
\addtolength{\arraycolsep}{-3pt}
\end{table}
In the SM the Wilson coefficients $C_7^{eff}(m_b)$ and $C_{10}(m_b)$, 
whose analytical expressions 
are given in \cite{R23,R24}, are
strictly real as can be read off from Table 2. 
In the leading logarithmic approximation, at the scale       
${\cal O}(\mu=m_b)$, we have                          
\bea                                                                         
C_7^{eff}(m_b) &=& -0.313~,\nnb \\             
C_{10}^{eff}(m_b) &=& -4.669~.\nnb                                           
\eea

\begin{table}[ht]
\renewcommand{\arraystretch}{1.5}
\addtolength{\arraycolsep}{3pt}
$$
\begin{array}{|c|c|c|c|c|c|c|c|c|c|}
\hline
C_{1} & C_{2} & C_{3} & C_{4} & C_{5} & C_{6}
& C_{7}^{eff} &
C_{9} & C_{10}^{eff} & C^{(0)} \\ \hline
-0.248 & 1.107 & 0.011 & -0.026 & 0.007 & -0.031 & -0.313 & 4.344 & -4.669
 & 0.362 \\ \hline  
\end{array}
$$
\caption{The numerical values of the Wilson coefficients at $\mu\sim m_{b}$
scale within the SM.}
\renewcommand{\arraystretch}{1}
\addtolength{\arraycolsep}{-3pt}
\end{table}
 
Although individual Wilson coefficients at $\mu \sim m_b$ level are all real
(see Table 2), the effective Wilson coefficient
$C_{9}^{eff}(m_{b},\hat s)$ has a finite phase, and in next--to--leading
order

\bea
\label{c9} 
C_9^{eff}(m_b,\hat s) = C_9(m_b)\left[1 + \frac{\alpha_s(\mu)}{\pi} \omega
(\hat s) \right]
+ Y_{SD}(m_b,\hat s) + Y_{LD}(m_B,s) ~,
\eea
where $C_9(m_b)=4.344$. Here $\omega \ga \hat s \dr$ represents the
${\cal{O}}(\alpha_{s})$ corrections
coming from one--gluon exchange in the matrix element of the corresponding
operator, whose explicit form can be found in \cite{R23}.
In (\ref{c9}) $Y_{SD}$ and $Y_{LD}$ represent, respectively, the short-- and
long--distance contributions of
the four--quark operators ${\cal{O}}_{i=1,\cdots,6}$ \cite{R23,R24}. Here
$Y_{SD}$ can be obtained
by a perturbative calculation  
\bea
Y_{SD}\ga m_{b}, \hat s \dr &=& g \ga \hat m_c,\hat s \dr
C^{(0)}
- \frac{1}{2} g \ga 1,\hat s \dr
\left[4 C_3 +4 C_4 + 3 C_5 + C_6 \right] \nnb \\
&-& \frac{1}{2} g \ga 0,\hat s \dr
\left[ C_3 + 3  C_4 \right]
+ \frac{2}{9} \left[ 3 C_3 + C_4 + 3 C_5 + C_6 \right] \nnb \\
&-& \lambda_u 
\left[ 3 C_1 + C_2 \right] \left[ g \ga 0,\hat s \dr -
g \ga \hat m_c,\hat s \dr \right]~,\nnb
\eea
where 
\bea
C^{(0)} &=& 3 C_1 + C_2 + 3 C_3 + C_4 + 3 C_5 + C_6~,\nnb \\ \nnb \\
\lambda_u &=& \frac{V_{ub} V_{ud}^\ast}{V_{tb} V_{td}^\ast}~, \nnb
\eea
and the loop function $g \ga m_q, s \dr$ stands for the loops
of quarks with mass $m_{q}$ at the dilepton invariant mass $s$.
This function develops absorptive parts for dilepton energies  
$s= 4 m_q^{2}$:
\bea
\lefteqn{
g \ga \hat m_q,\hat s \dr = - \frac{8}{9} \ln \hat m_q +
\frac{8}{27} + \frac{4}{9} y_q -
\frac{2}{9} \ga 2 + y_q \dr \sqrt{\vel 1 - y_q \ver}} \nnb \\
&&\times \Bigg[ \Theta(1 - y_q)
\ga \ln \frac{1  + \sqrt{1 - y_q}}{1  -  \sqrt{1 - y_q}} - i \pi \dr
+ \Theta(y_q - 1) \, 2 \, \arctan \frac{1}{\sqrt{y_q - 1}} \Bigg], \nnb
\eea
where  $\hat m_q= m_{q}/m_{b}$ and $y_q=4 \hat m_q^2/\hat s$. 
In addition to these perturbative contributions, the $\bar{c}c$ loops
can excite low--lying charmonium states $\psi(1s), \cdots, \psi(6s)$ 
whose contributions are represented by $Y_{LD}$ \cite{R25}: 
\bea
Y_{LD}\ga m_{b}, \hat s \dr&=& \frac{3}{\alpha^2}
\Bigg[ - \frac{V_{cf}^* V_{cb}}{V_{tf}^* V_{tb}} \,C^{(0)} -
\frac{V_{uf}^* V_{ub}}{V_{tf}^* V_{tb}}
\ga 3 C_3 + C_4 + 3 C_5 + C_6 \dr \Bigg] \nnb \\
&\times& \sum_{V_i = \psi \ga 1 s \dr, \cdots, \psi \ga 6 s \dr}
\ds{\frac{ \pi \kappa_{i} \Gamma \ga V_i \rar \ell^+ \ell^- \dr M_{V_i} }
{\ga M_{V_i}^2 - \hat s m_b^2 - i M_{V_i} \Gamma_{V_i} \dr }}~, \nnb
\eea
where $\kappa_i$ are the Fudge factors (see for example \cite{R7}).

Let us first study the dependence  of the asymmetry parameter
$\alpha_{\theta_P}$ on the new Wilson coefficients. Note that in further
analysis, only short distance contributions are taken into account and
integration over $q^2$ is performed in the full physical region $4 m_\ell^2
\le q^2 \le (m_B-m_V^2)^2$. We assumed that all new Wilson coefficients
$C_X$ are real, i.e., we do not introduce any new phase in addition to the
one present in the SM.

In Figs. (1) and (2), we present the dependence of $\alpha_{\theta_P}$ on
the new Wilson coefficients, for the $B \rar \pi^+ \pi^- e^+ e^-$ and
$B \rar \pi^+ \pi^- \tau^+ \tau^-$ decays, respectively. 
Here and in all of the following figures, zero value of new Wilson
coefficients $C_X$ correspond to the SM prediction. 
In the case of 
$B \rar \pi^+ \pi^- e^+ e^-$ decay the asymmetry parameter $\alpha_{\theta_P}$ is
more sensitive to $C_{LL}^{tot}$ and $C_{RL}$, while for the $B \rar \pi^+ \pi^-
\tau^+ \tau^-$ decay it depends strongly on $C_{RL}$. These dependencies
can be explained as follows. For the $B \rar \pi^+ \pi^- e^+ e^-$ decay, if
the terms proportional to electron mass are neglected, it easily be seen from 
Eq. (\ref{alp}) that
\bea
\alpha_{\theta_P} = \frac{\widetilde N_2}{\widetilde N_6} - 1 ~.\nnb
\eea
In the limit $v \rar 1$ we get 
\bea
\eqalign{
\widetilde N_2 &\simeq \vel q^2 \ver \vel 2 \ga C_9^{eff} + C_{!0} + C_{LR}^{tot}
\dr H_0 - 4 C_7^{eff} \frac{m_b}{q^2} {\cal H}_0 + 
C_{RR} h_0  \ver^2 \cr
&+\vel q^2 \ver \vel 2 \ga C_9^{eff} - C_{!0} + C_{LL}^{tot}
\dr H_0 - 4 C_7^{eff} \frac{m_b}{q^2} {\cal H}_0 + 
C_{RL
} h_0  \ver^2~.} \cr
\eea 
In the SM in the large dilepton mass region, say about $q^2 \simeq 5~GeV^2$,
$C_9^{eff} + C_{10} \simeq 0.4$ and Re$\left[C_9^{eff} - C_{10}\right]
\simeq 9.5$. It follows then that the interference terms between 
$C_9^{eff} - C_{10}$ and $C_{LL}~(C_{RL})$ are dominant and hence
contributions coming from $C_{LL}~(C_{RL})$ are large. These figures
illustrates that the contributions of $C_{LL}$ and $C_{RL}$ to 
$\alpha_{\theta_P}$ is positive for $C_{LL}>0$ and $C_{RL}<0$, and negative for 
$C_{LL}<0$ and $C_{RL}>0$. It should be noted that the asymmetry parameter
$\alpha_{\theta_P}$ can get only positive or negative values for the case 
$C_{LL}\ne 0$ and $C_{RL}\ne 0$, while it is always positive for all other
choices of the Wilson coefficients, as is the case in the SM. For this
reason determination of the sign of $\alpha_{\theta_P}$ can serve as an
efficient tool for establishing new physics. 

In the $B \rar \pi^+ \pi^- \tau^+ \tau^-$ process however, the situation is
slightly different compared to that of the $B \rar \pi^+ \pi^- e^+ e^-$ 
transition. Largest contribution in this case comes from $C_{RL}$ and
contributions from all other Wilson coefficients are comparable to one
another. This observation can be attributed to mass of the $\tau$ lepton,
for which $\alpha_{\theta_P}$ is positive for the choice of each individual 
new Wilson coefficients. 

Depicted in Figs. (3), (4) and Figs. (5), (6) are the dependencies of the
asymmetry parameters $\alpha_\theta$ and $\beta_\theta$ on the new Wilson
coefficients $C_X$, for the $B \rar \pi^+ \pi^- e^+ e^-$ and $B \rar \pi^+
\pi^- \tau^+ \tau^-$ decays, respectively. Figs. (3) and (5) depict that the
asymmetry parameter $\alpha_\theta$ for the $e^+ e^-$ channel depends
strongly on the new Wilson coefficient $C_{RL}$, while it displays similar
behavior for all new Wilson coefficients for the $\tau^+ \tau^-$ case. 
We observe that the asymmetry parameter $\beta_\theta$ depends strongly on
$C_{RL}$ in the $e^+ e^-$ channel and on $C_{RL}$ and $C_{RR}$ in the
$\tau^+ \tau^-$ channel. It is interesting that $\beta_\theta$ changes its
sign when $C_{RL} > 2$ in the $e^+ e^-$ channel, while it is negative for
all values of $C_{RL}$ or $C_{RR}$ in the $\tau^+ \tau^-$ channel. Therefore
determination of the sign of $\beta_\theta$ is useful in looking for new
physics. 

In Figs. (7) and (8) we present the dependence of the asymmetry parameter
$\beta_\varphi$ for the $B \rar \pi^+ \pi^- e^+ e^-$ and 
$B \rar \pi^+ \pi^-  \tau^+ \tau^-$ decays, respectively. From these figures
one notices that this asymmetry parameter is quite sensitive to the
variation in $C_{RL}$ and changes its sign at $C_{RR}>2$ for both channels. 
Our investigation of the dependence of the
asymmetry parameter $\alpha_\varphi$ on the new Wilson coefficients shows
that for the range $-4 < C_X < 4$, $\alpha_\varphi$ varies between 
$-6\times 10^{-3}$ to $6\times 10^{-3}$ for the 
$B \rar \pi^+ \pi^- e^+ e^-$ and $-5.0\times 10^{-3}$ to $2.5\times 10^{-3}$ 
for the $B \rar \pi^+ \pi^- \tau^+ \tau^-$ decays, repectively.
Therefore detection of the dependence of the asymmetry parameter
$\alpha_\varphi$ on the new Wilsom coefficients is quite hard from
experimental point of view. 

The asymmetry parameter $\widetilde A_1$ shows strong dependence on 
$C_{RL}$ and $C_{LL}$ for the $e^+ e^-$ channel, as depicted in Fig. (9),
whose contributions are 
dominant compared to the other Wilson coefficients. For the $\tau^+ \tau^-$
channel contributions of $C_{RL}$ and $C_{RR}$ become dominant, as can be
seen in Fig. (10).  
The asymmetry parameter $\widetilde A_2$ varies considerably
for the $e^+ e^-$ channel,
in relation to the variations occurring in $C_{RL}$, while this behavior is
switched to the Wilson coefficient $C_{LRRL}$ for the $\tau^+ \tau^-$ case
which are presented in Figs. (11) and (12).

Presented in Figs. (13) and (14) are the dependence of the asymmetry
parameter $A_\varphi$ on new Wilson coefficients for the $e^+ e^- \pi^+
\pi^-$ and $\tau^+ \tau^- \pi^+ \pi^-$ decays, respectively. In both cases
the asymmetry parameter $A_\varphi$ shows strong dependence on $C_{RL}$.   

Finally, in Figs. (15) and (16) we present the dependence of the averaged 
(i.e., integrated over $q^2$ in the full physical region) CP asymmetry on the
Wilson coefficients for the $B \rar \pi^+ \pi^- e^+ e^-$ and $B \rar \pi^+
\pi^-  \tau^+ \tau^-$ decays, respectively. We observe that 
$\lla A_{CP} \rra$ is strongly dependent on $C_{RL}$ and $C_{RR}$ for the 
$e^+ e^-$ channel, while this dependence is switched to $C_{LL}$ for the
$\tau^+ \tau^-$ channel. One can easily read from this figures that 
$\lla A_{CP}\rra  > -2.0 \times 10^{-2}$ in the $e^+ e^-$ channel, 
for the values $C_{RR} \le -2$ and 
$\vel \lla A_{CP}\rra \ver  > 1.7 \times 10^{-2}$ 
when $C_{LL} \le -2$.

As the final concluding remark, we presented in this work the model
independent analysis of the exclusive $B \rar \pi^+ \pi^- \ell^+ \ell^-$
$(\ell = e,~\tau)$ decay is presented. In particular, the sensitivity to 
the new Wilson coefficients of the experimentally measurable asymmetries and
CP violating asymmetry are systematically analyzed. The main result of the
present study is that different asymmetry parameters show strong dependence on
different new Wilson coefficients. Therefore a combined analysis of the
different asymmetries and CP violating asymmetry can give unambiguous
information about the existence of new physics beyond the SM and especially
about various new Wilson coefficients.

\newpage
\section*{Figure captions}
{\bf Fig. 1} The dependence of the asymmetry parameter $\alpha_{\theta_P}$
on the new Wilson coefficients for the $B \rar \pi^+ \pi^- e^+ e^-$ 
decay. \\ \\
{\bf Fig. 2} The same as in Fig. (1), but for the 
$B \rar \pi^+ \pi^- \tau^+ \tau^-$ decay. \\ \\ 
{\bf Fig. 3} The same as in Fig. (1), but for the asymmetry parameter
$\alpha_\theta$.\\ \\
{\bf Fig. 4} The same as in Fig. (1), but for the asymmetry parameter
$\beta_\theta$.\\ \\
{\bf Fig. 5} The same as in Fig. (3), but for the 
$B \rar \pi^+ \pi^- \tau^+ \tau^-$ decay. \\ \\
{\bf Fig. 6} The same as in Fig. (4), but for the 
$B \rar \pi^+ \pi^- \tau^+ \tau^-$ decay. \\ \\
{\bf Fig. 7} The same as in Fig. (1), but for the asymmetry parameter
$\beta_\varphi$.\\ \\
{\bf Fig. 8} The same as in Fig. (7), but for the 
$B \rar \pi^+ \pi^- \tau^+ \tau^-$ decay. \\ \\
{\bf Fig. 9} The same as in Fig. (1), but for the asymmetry parameter
$\widetilde A_1$. \\ \\
{\bf Fig. 10} The same as in Fig. (9), but for the 
$B \rar \pi^+ \pi^- \tau^+ \tau^-$ decay. \\ \\
{\bf Fig. 11} The same as in Fig. (9), but for the asymmetry parameter
$\widetilde A_2$.\\ \\
{\bf Fig. 12} The same as in Fig. (11), but for the 
$B \rar \pi^+ \pi^- \tau^+ \tau^-$ decay. \\ \\
{\bf Fig. 13} The same as in Fig. (9), but for the asymmetry parameter
$\widetilde A_\varphi$.\\ \\
{\bf Fig. 14} The same as in Fig. (13), but for the
$B \rar \pi^+ \pi^- \tau^+ \tau^-$ decay. \\ \\
{\bf Fig. 15} The dependence of the averaged CP asymmetry 
on the new Wilson coefficients for the $B \rar \pi^+ \pi^- e^+ e^-$
decay. \\ \\
{\bf Fig. 16} The same as in Fig. (15), but for the 
$B \rar \pi^+ \pi^- \tau^+ \tau^-$ decay. \\ \\

\newpage

\begin{figure}
\vskip 1cm
    \includegraphics{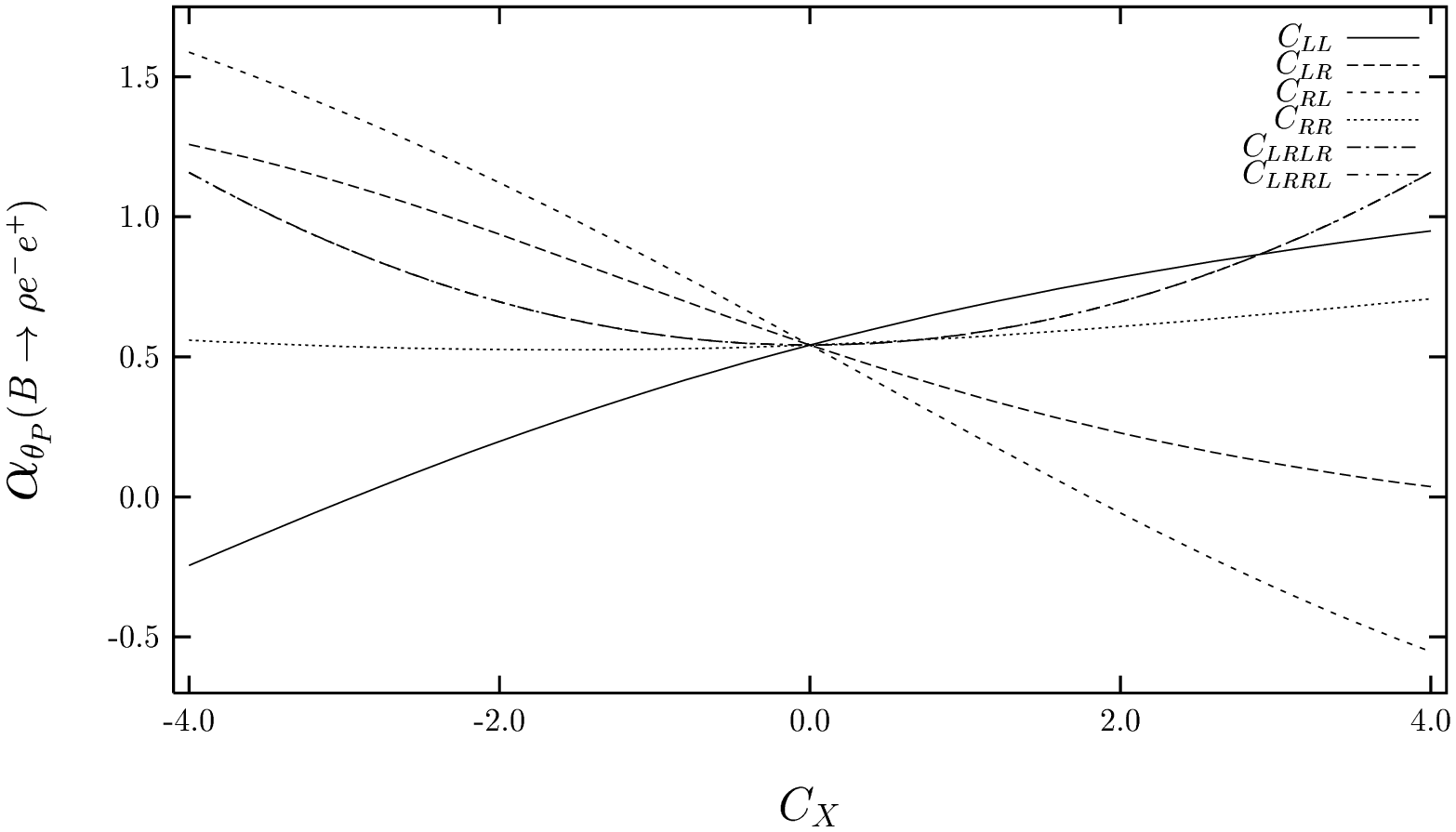}
\vskip 8.1cm
\caption{}
\end{figure}

\begin{figure}
\vskip 1.5 cm
    \includegraphics{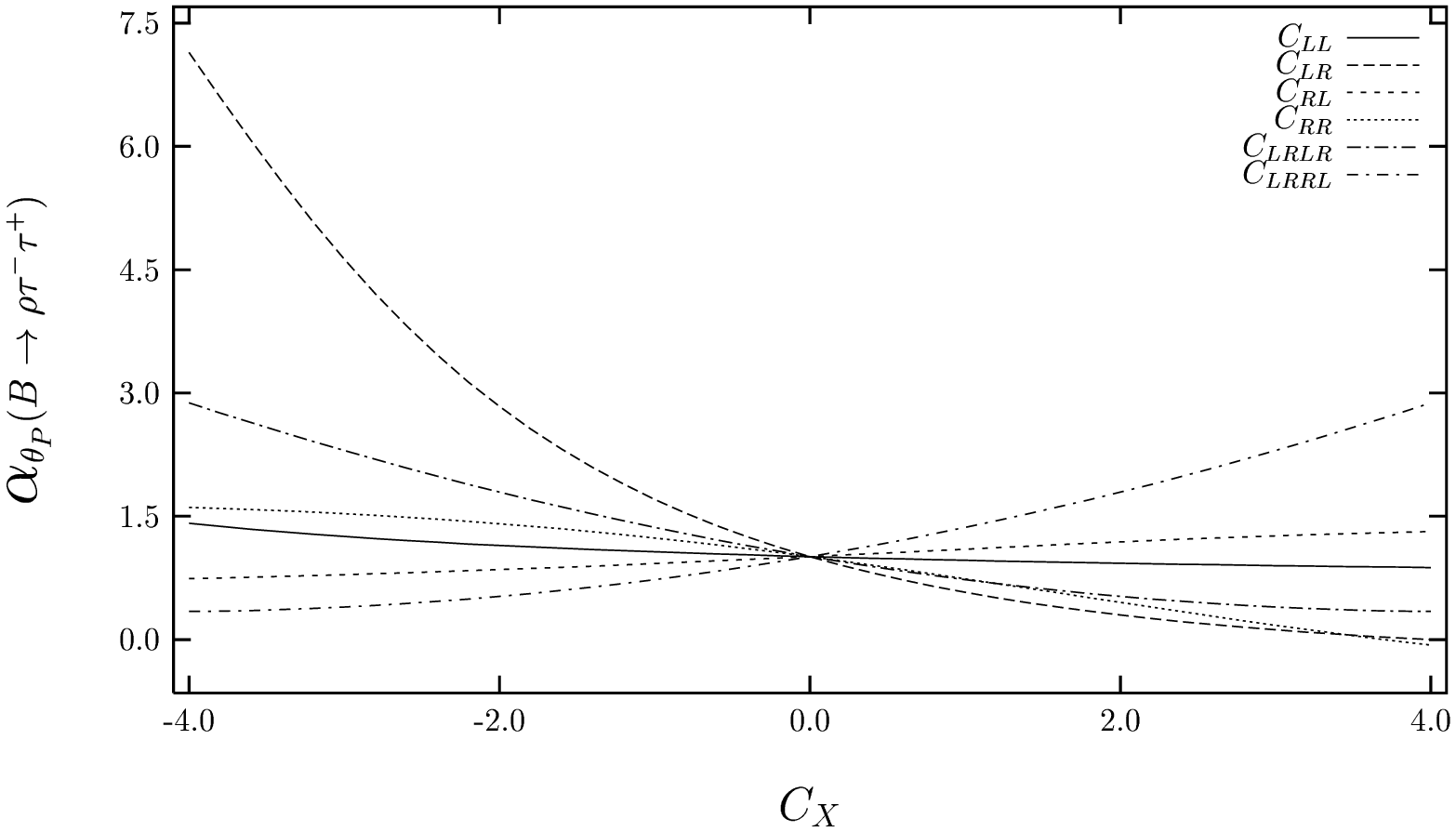}
\vskip 9. cm
\caption{}
\end{figure}

\begin{figure}
\vskip 1.5 cm
    \includegraphics{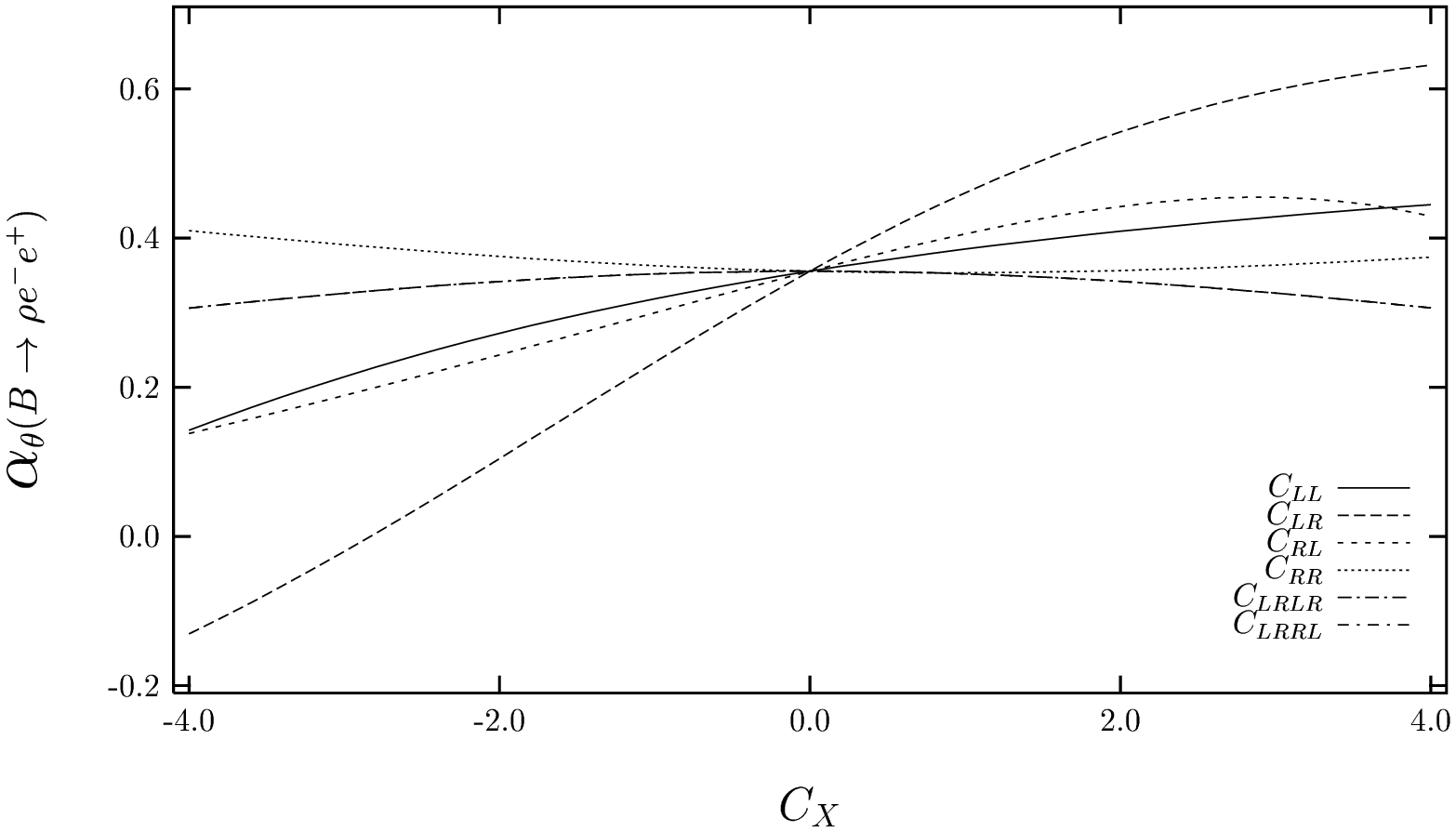}
\vskip 9. cm
\caption{}
\end{figure}

\begin{figure}
\vskip 1cm
    \includegraphics{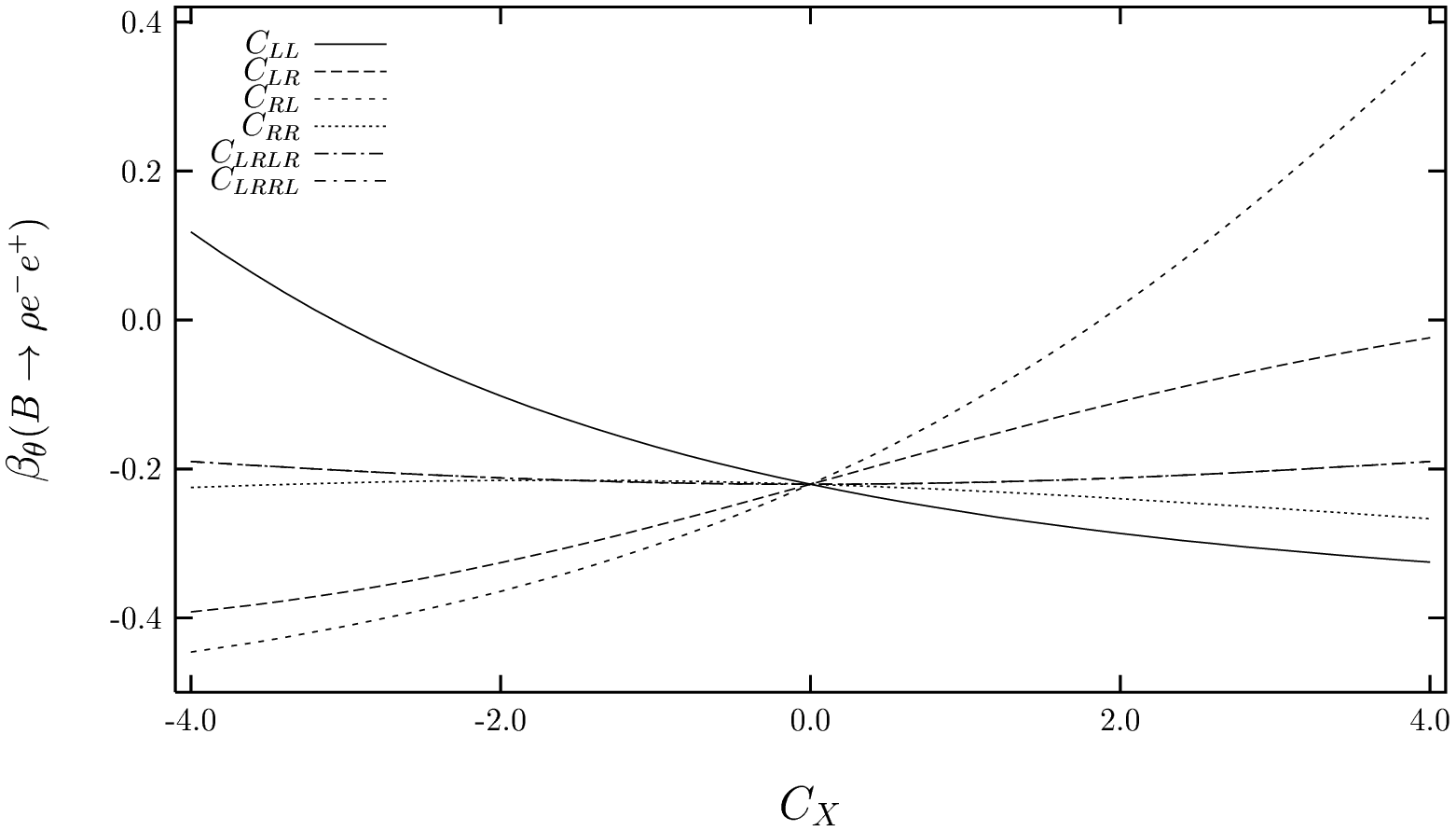}
\vskip 8.1cm
\caption{}
\end{figure}

\begin{figure}
\vskip 1.5 cm
    \includegraphics{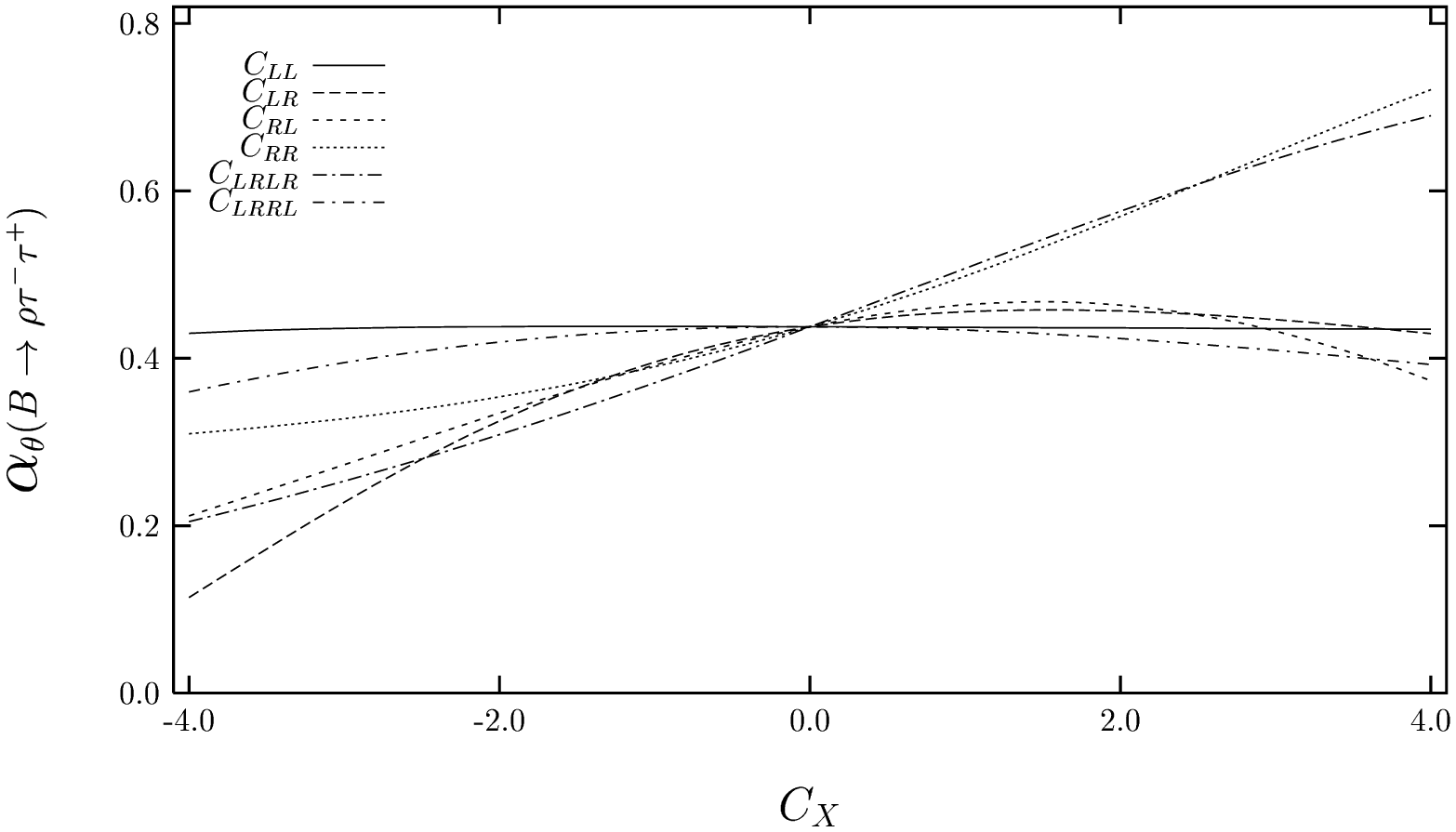}
\vskip 9. cm
\caption{}
\end{figure}

\begin{figure}
\vskip 1cm
    \includegraphics{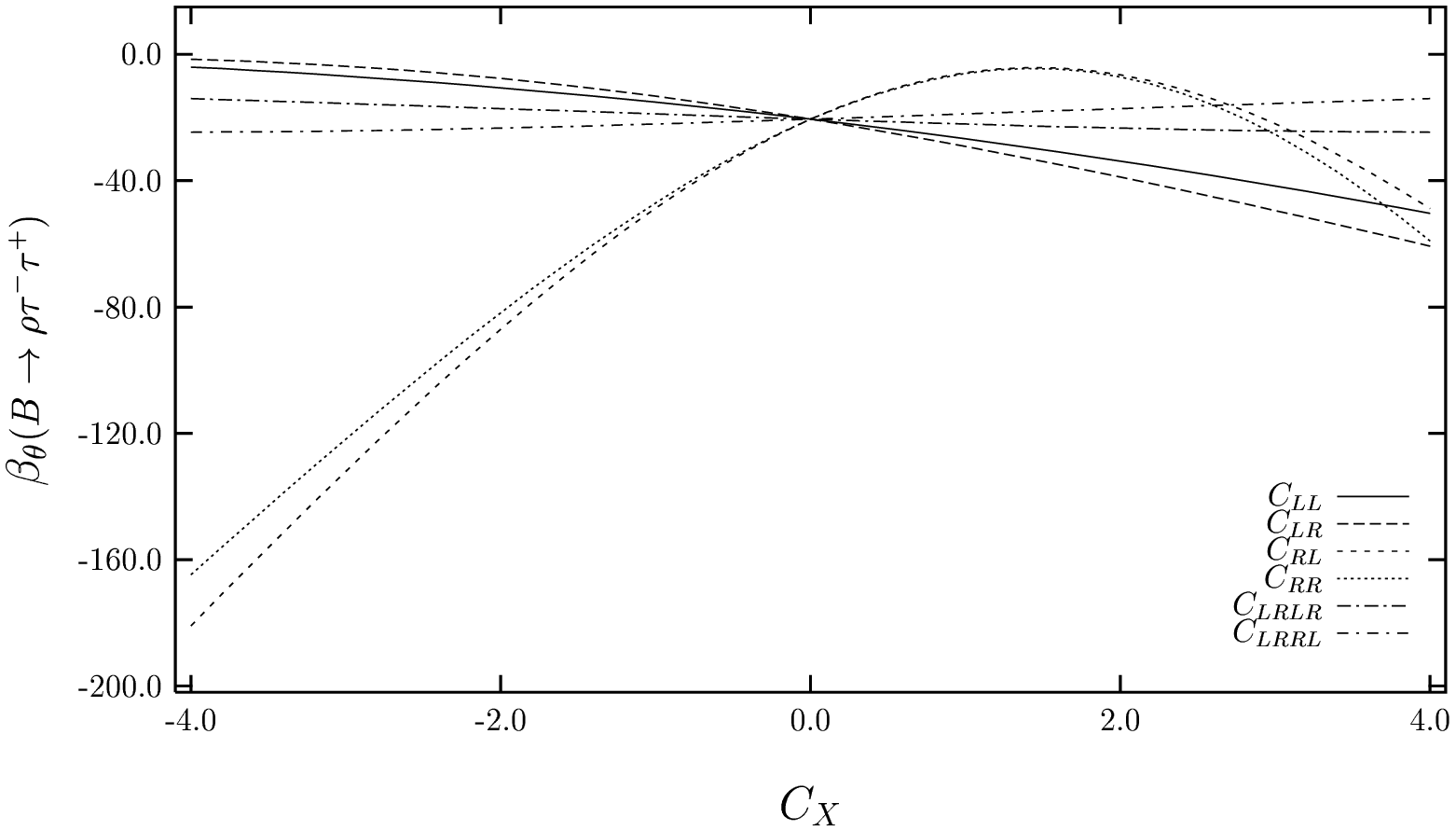}
\vskip 8.1cm
\caption{}
\end{figure}

\begin{figure}
\vskip 1.5 cm
    \includegraphics{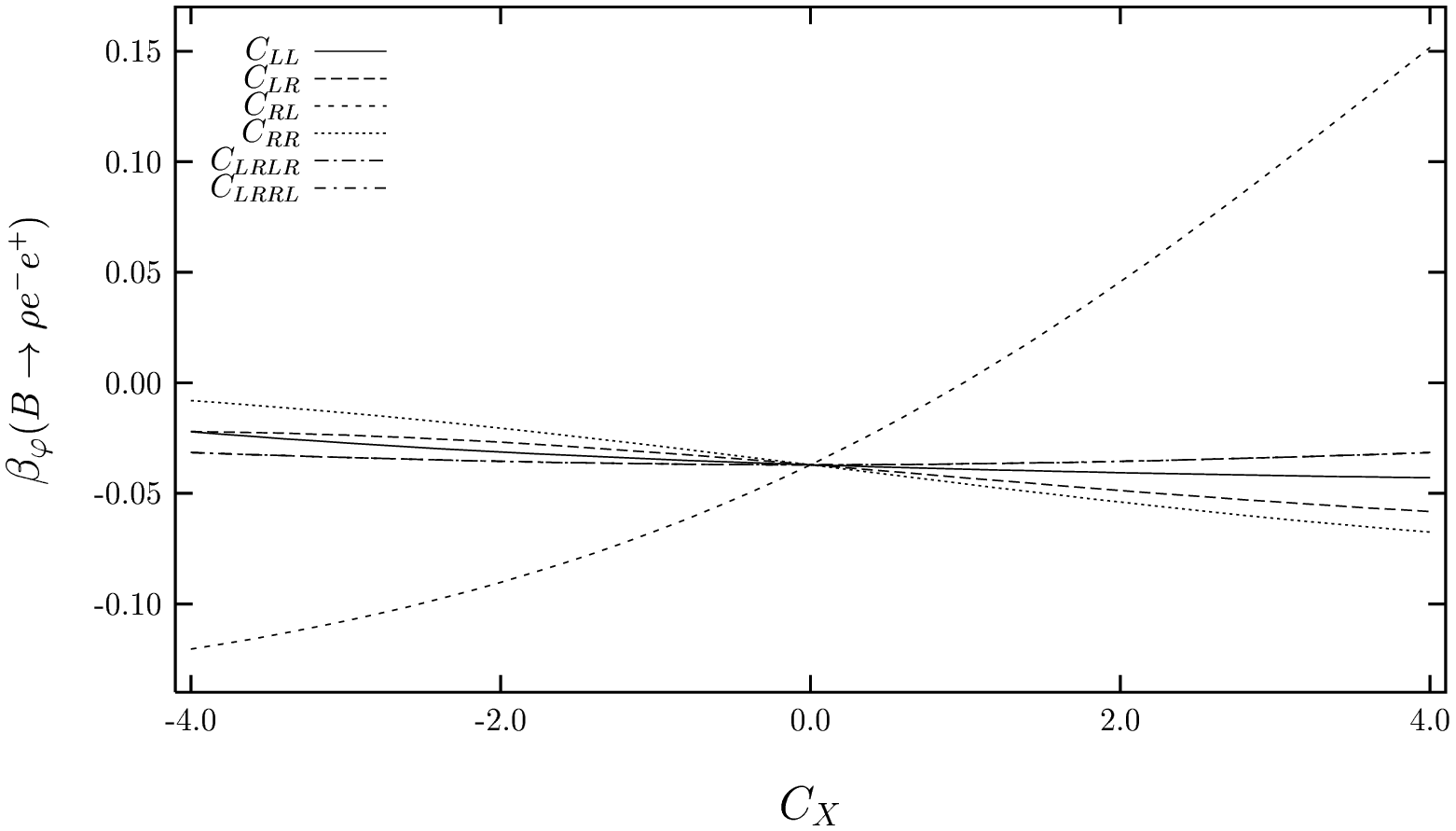}
\vskip 9. cm
\caption{}
\end{figure}

\begin{figure}
\vskip 1cm
    \includegraphics{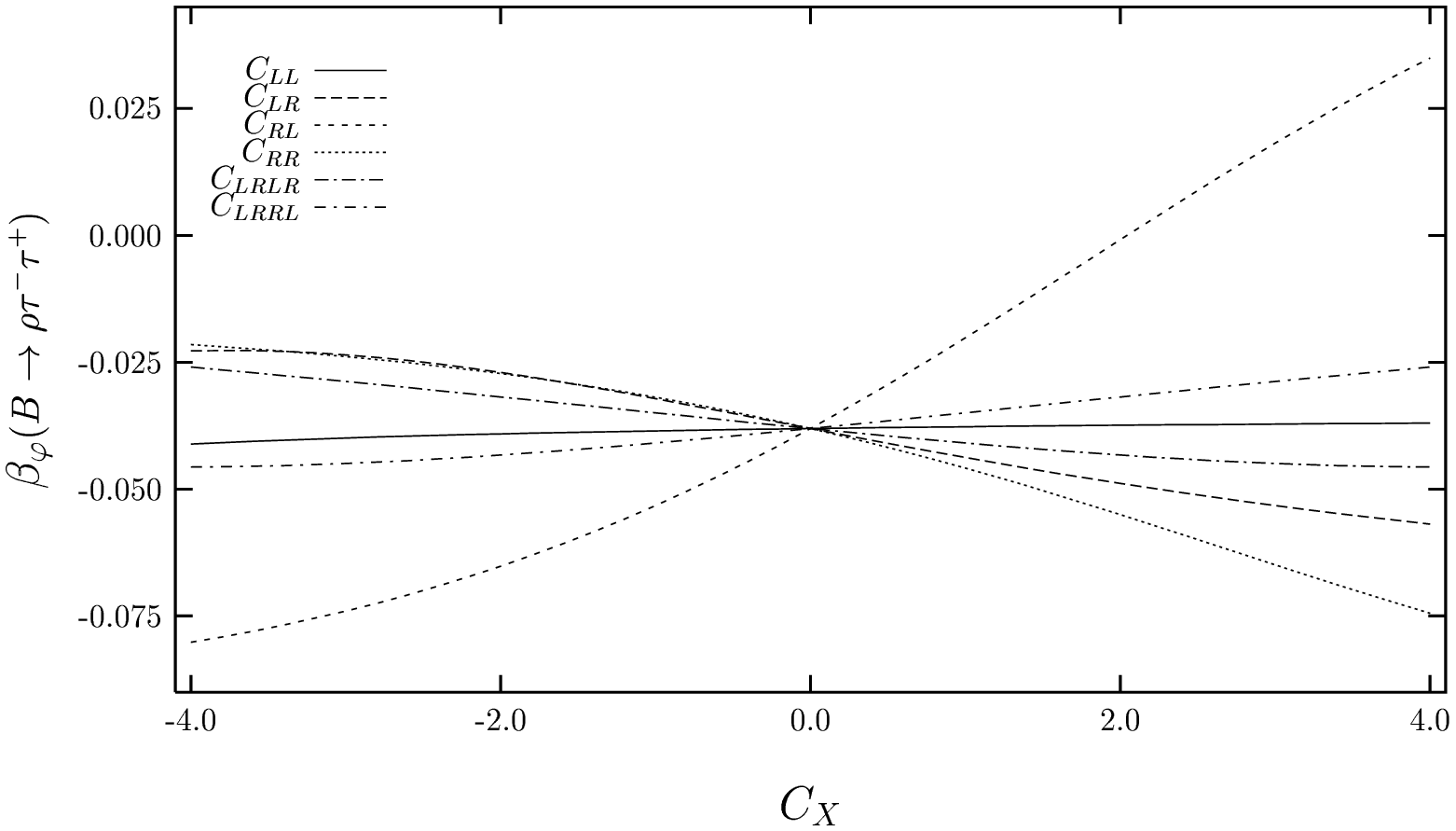}
\vskip 8.1cm
\caption{}
\end{figure}

\begin{figure}
\vskip 1.5 cm
    \includegraphics{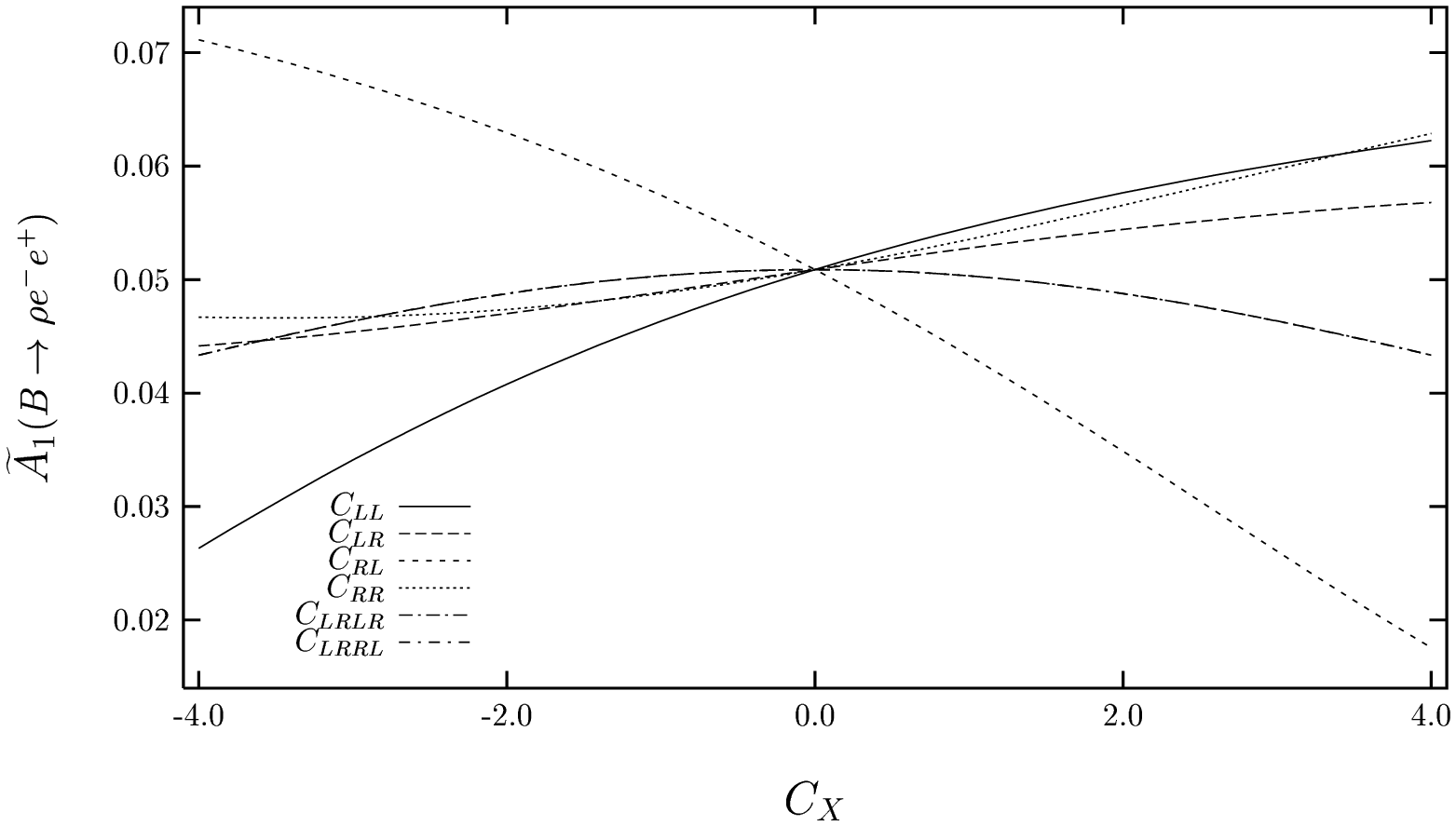}
\vskip 9. cm
\caption{}
\end{figure}

\begin{figure}
\vskip 1cm
    \includegraphics{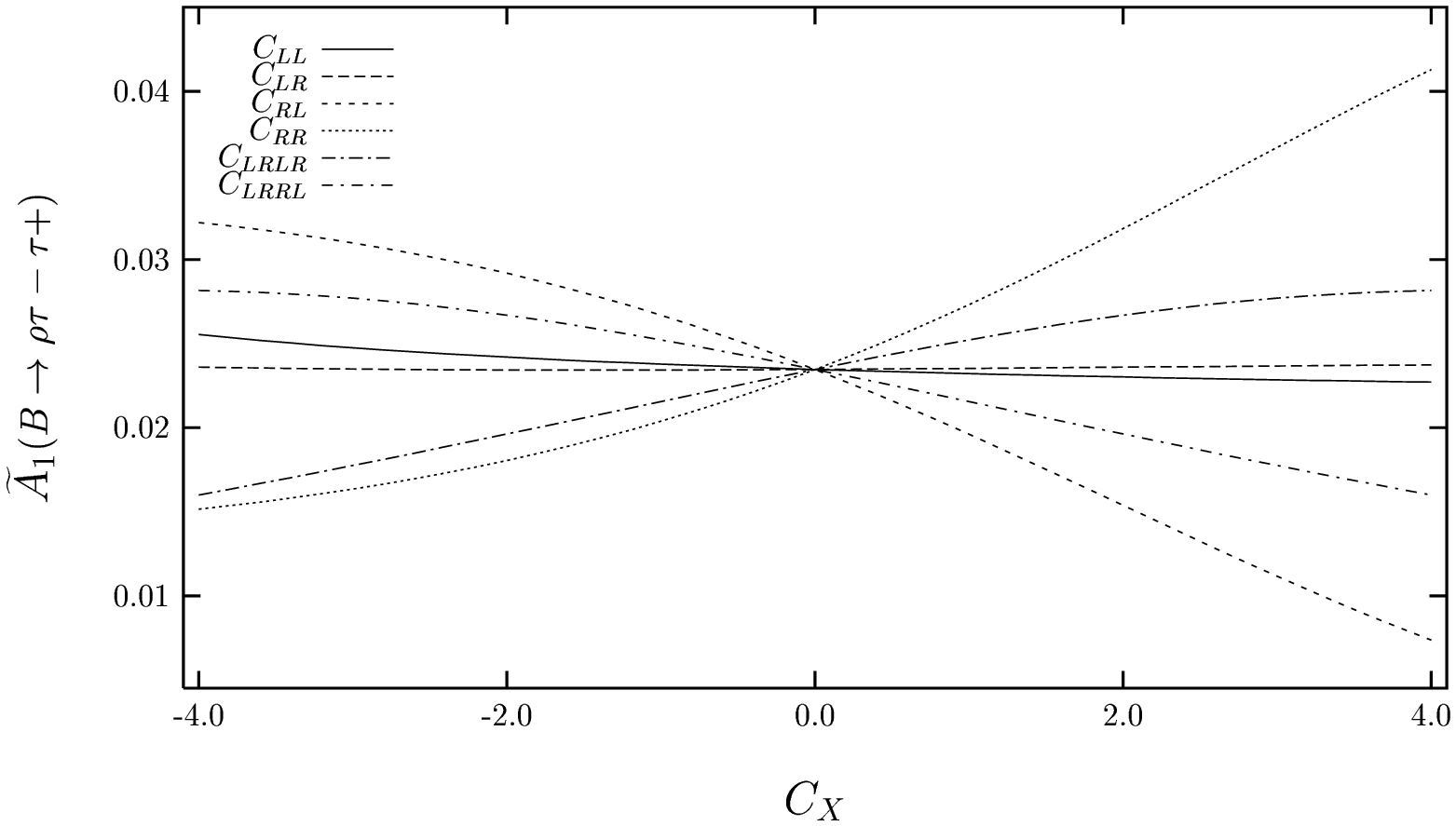}
\vskip 8.1cm
\caption{}
\end{figure}

\begin{figure}
\vskip 1.5 cm
    \includegraphics{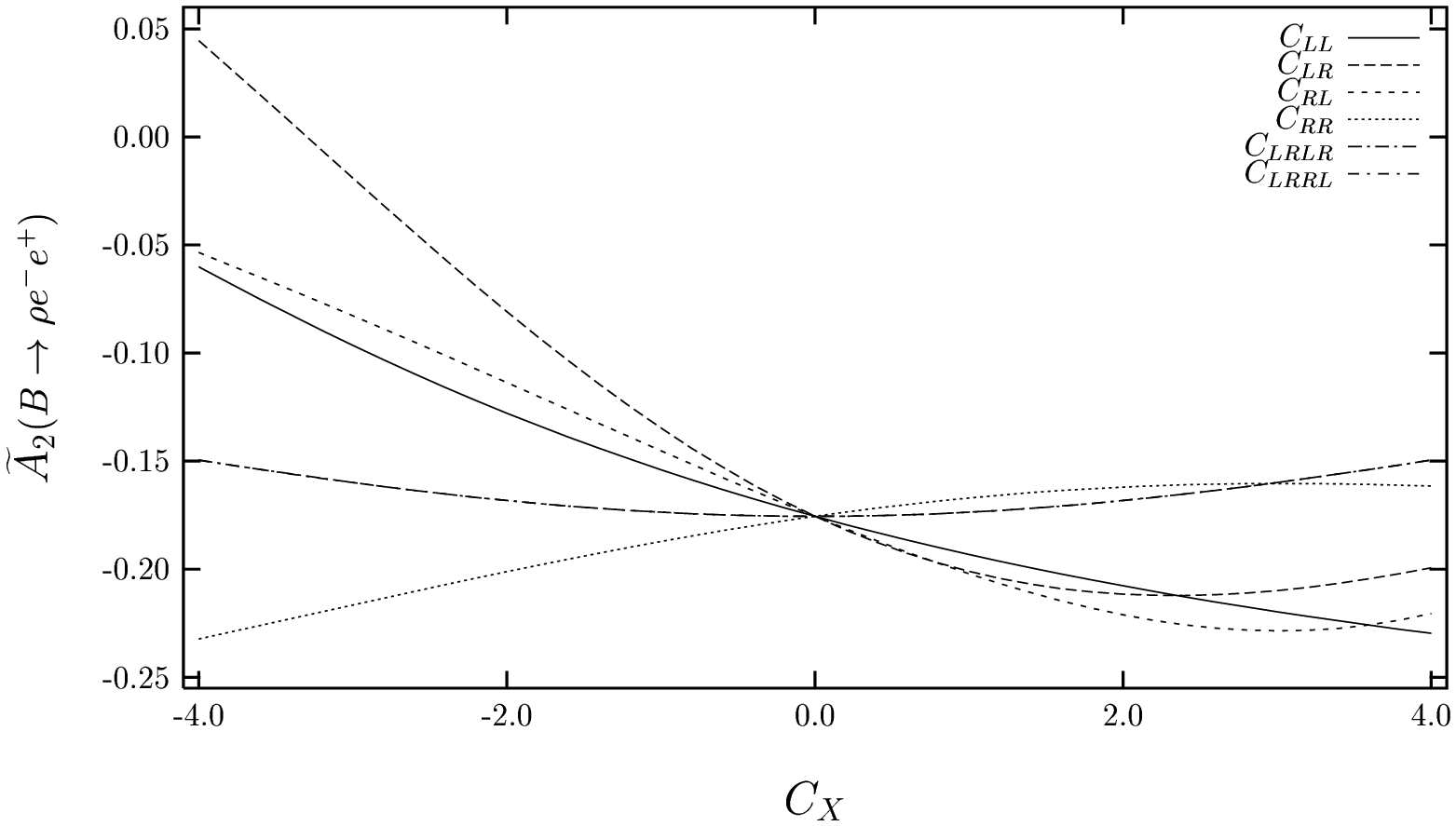}
\vskip 9. cm
\caption{}
\end{figure}

\begin{figure}
\vskip 1.5 cm
    \includegraphics{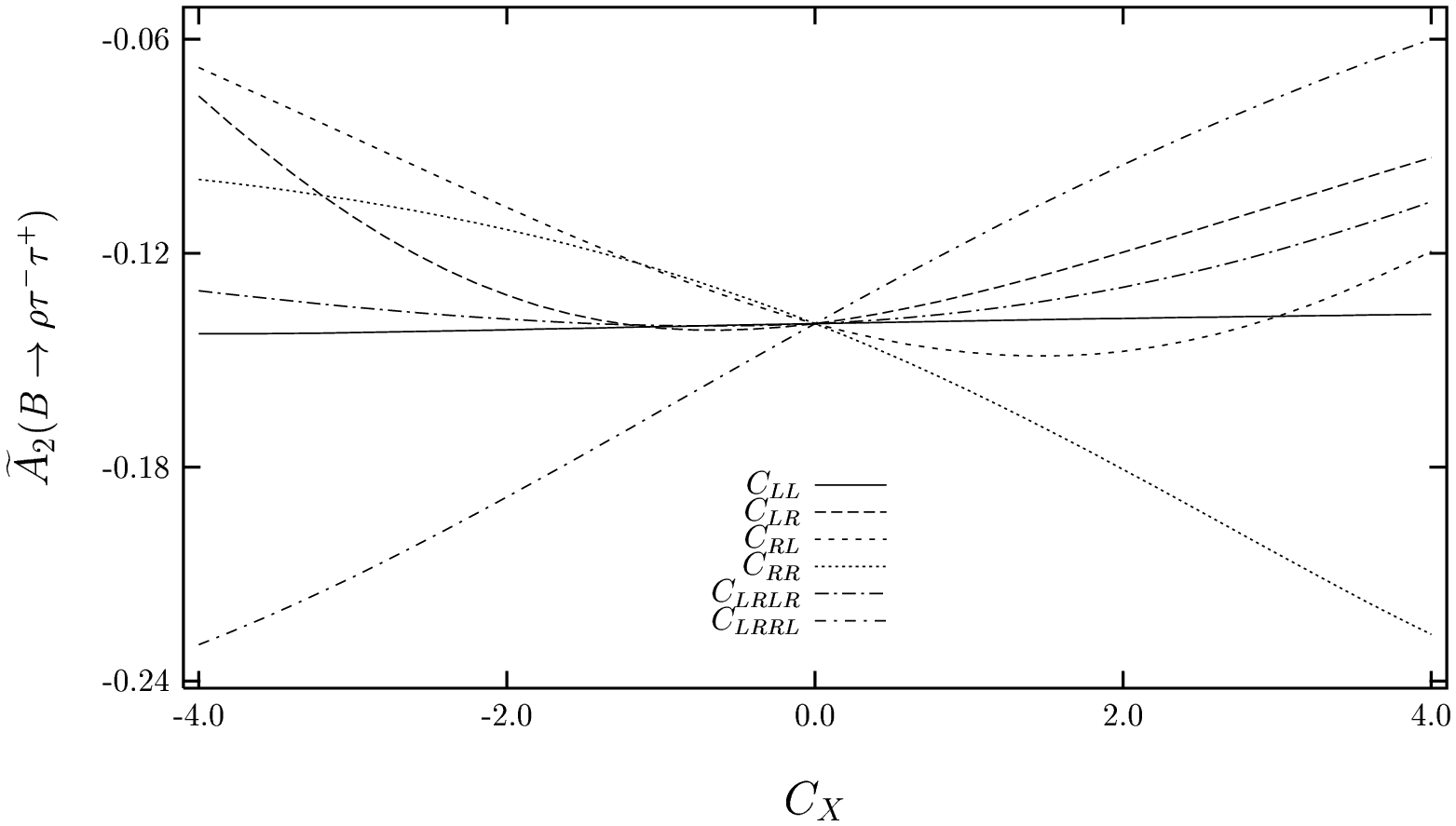}
\vskip 9. cm
\caption{}
\end{figure}

\begin{figure}
\vskip 1cm
    \includegraphics{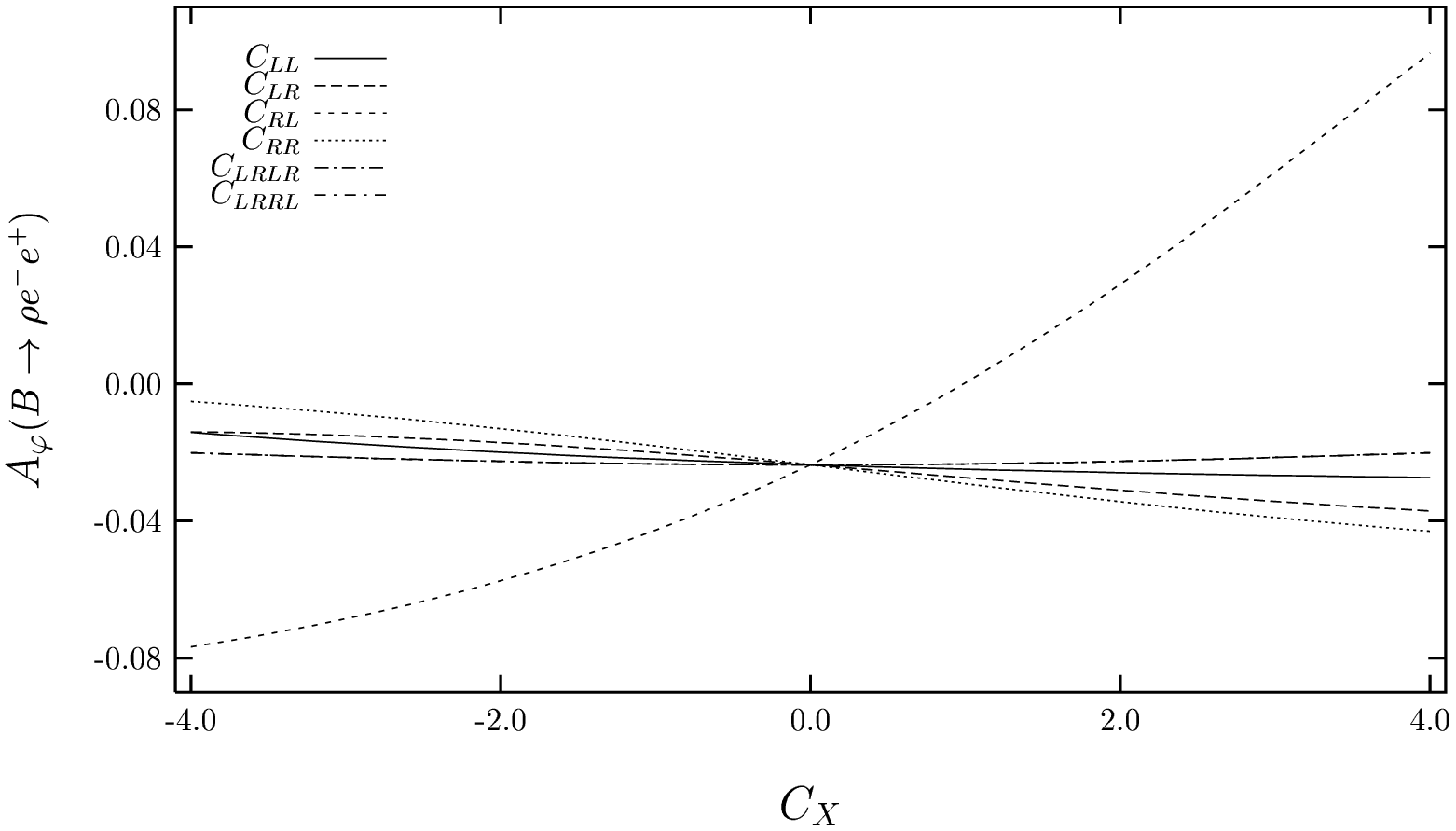}
\vskip 8.1cm
\caption{}
\end{figure}

\begin{figure}
\vskip 1.5 cm
    \includegraphics{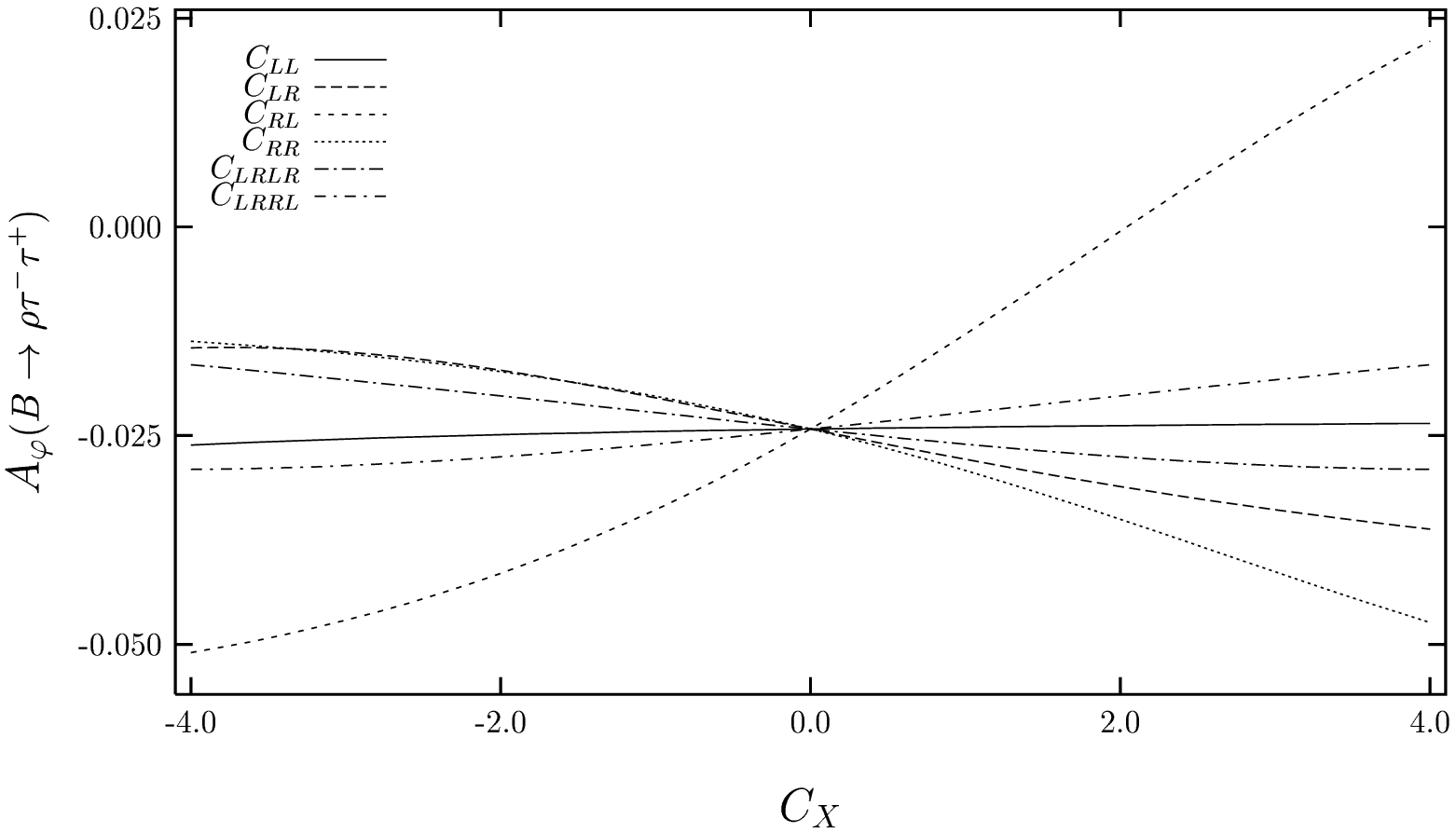}
\vskip 9. cm
\caption{}
\end{figure}

\begin{figure}
\vskip 1cm
    \includegraphics{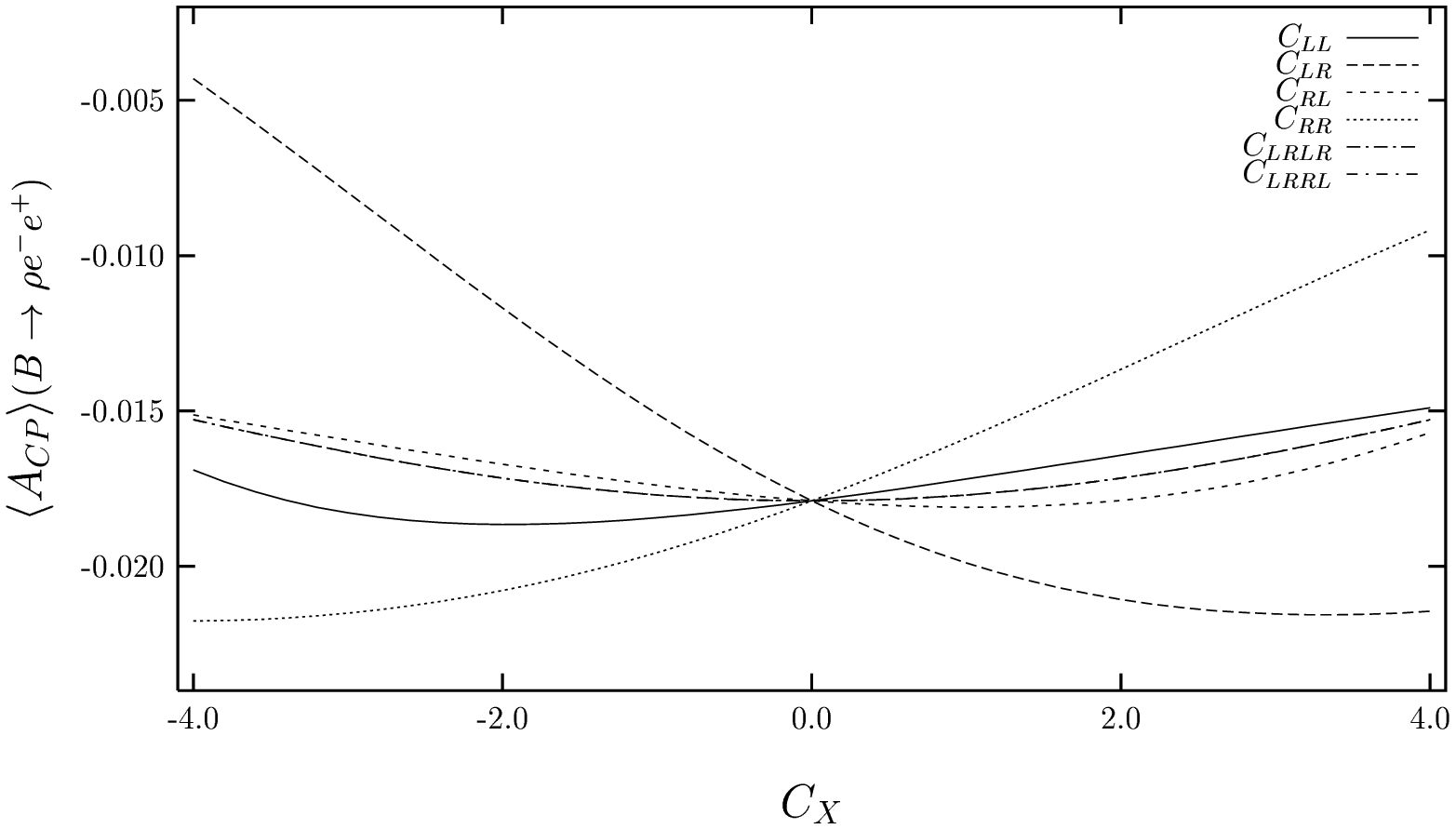}
\vskip 8.1cm
\caption{}
\end{figure}

\begin{figure}
\vskip 1.5 cm
    \includegraphics{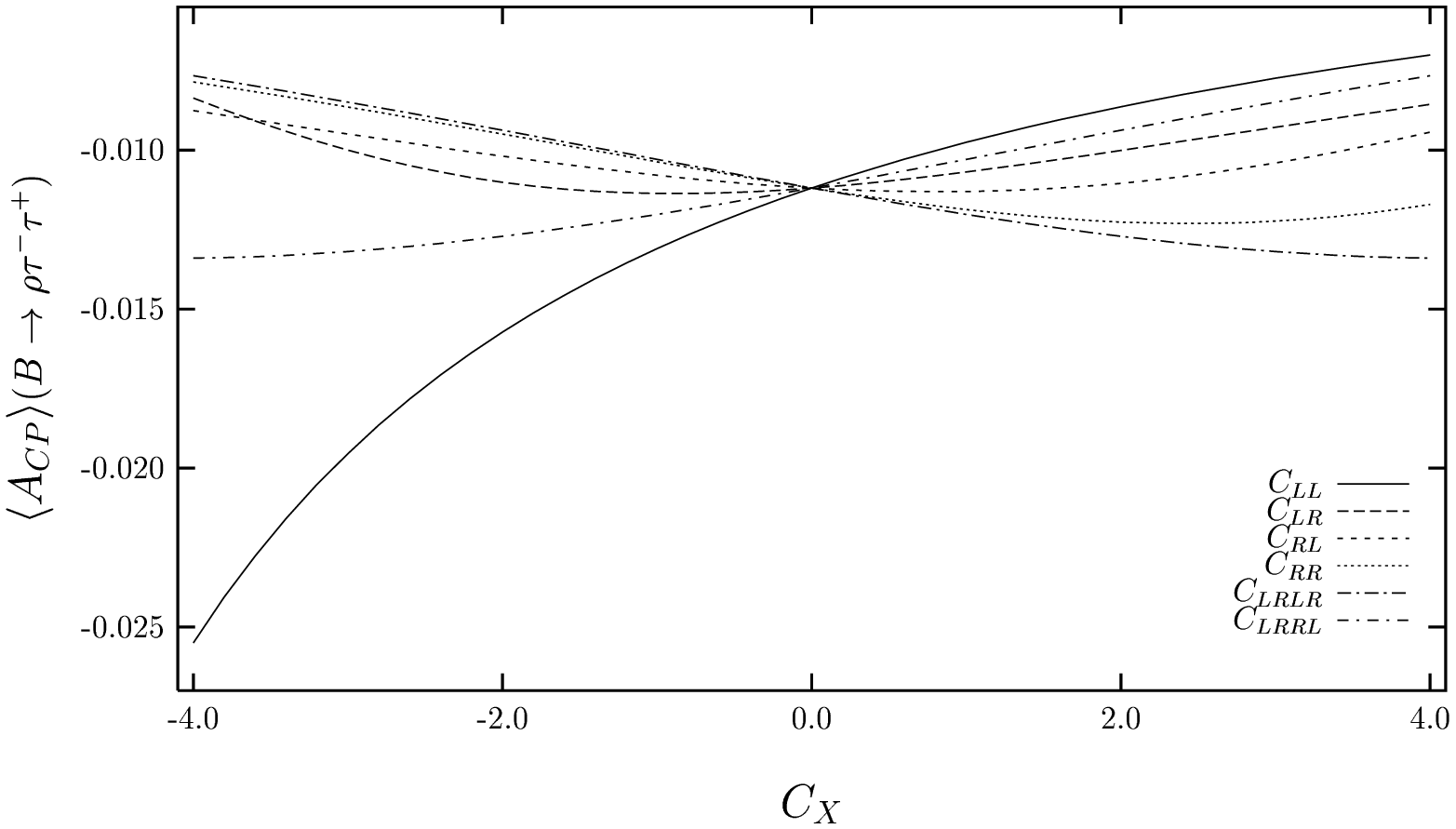}
\vskip 9. cm
\caption{}
\end{figure}

\end{document}